\documentclass[preprint,sort&compress]{elsarticle}
\usepackage[left=2.5cm,top=3.5cm,right=2.5cm,bottom=2cm,nohead]{geometry}

\usepackage[english]{babel}  
\usepackage[utf8]{inputenc}
\usepackage{type1ec}
\usepackage[T1]{fontenc}
\usepackage{multirow}
\usepackage{graphicx}
\usepackage{enumitem}
\usepackage[all]{nowidow}

\usepackage{amsmath, amsthm, amssymb}
\usepackage{algpseudocode}
\usepackage{algorithm}
\usepackage{multicol}
\usepackage{bm}
\usepackage{multirow}
\usepackage{listings}
\usepackage{color}
\usepackage[usenames,dvipsnames]{xcolor}
\usepackage{soul}
\usepackage{url}
\usepackage[lighttt]{lmodern}
\usepackage{setspace}
\usepackage{booktabs}
\usepackage{multirow}
\usepackage{caption}
\usepackage{subcaption}
\usepackage{tablefootnote}
\usepackage{scrextend}
\usepackage{tcolorbox}

\newcommand{\aspas}[1]{{``#1''}}

\newcommand{\sbarf}[2]{{\color{darkgray}\rule{\dimexpr 1cm * #1 / #2}{6pt}\color{lightgray}\rule{\dimexpr 1cm * (#2 - #1) / #2}{6pt}}}

\newcommand{\specialcell}[2][c]{%
  \begin{tabular}[#1]{@{}c@{}}#2\end{tabular}}

\newcommand{\mypm}{\mathbin{\mathpalette\@mypm\relax}}

\begin{document}

\begin{frontmatter}

\title{What's in a GitHub Star? \\
Understanding Repository Starring Practices in a Social Coding Platform}

\author{Hudson Borges}
\ead{hsborges@dcc.ufmg.br}
\address{Department of Computer Science, UFMG, Brazil}

\author{Marco Tulio Valente}
\ead{mtov@dcc.ufmg.br}
\address{Department of Computer Science, UFMG, Brazil}

\begin{abstract}
  %!TEX root = jss.tex

\noindent Besides a git-based version control system, GitHub integrates several social coding features. Particularly, GitHub users can \emph{star} a repository, presumably to manifest interest or satisfaction with an open source project. However, the real and practical meaning of {\em starring a project} was never the subject of an in-depth and well-founded empirical investigation.
Therefore, we provide in this paper a throughout study on the meaning, characteristics, and dynamic growth of GitHub stars.
First, by surveying 791 developers, we report that three out of four developers consider the number of stars before using or contributing to a GitHub project.
Then, we report a quantitative analysis on the characteristics of the top-5,000 most starred GitHub repositories.
We propose four patterns to describe stars growth, which are derived after clustering the time series representing the number of stars of the studied repositories; we also reveal the perception of 115 developers about these growth patterns. To conclude, we provide a list of recommendations to open source project managers (e.g.,~on the importance of social media promotion) and to GitHub users and Software Engineering researchers (e.g., on the risks faced when selecting projects by GitHub stars). 

\end{abstract}


\end{frontmatter}

% !TEX root = jss.tex
\section{Introduction}

GitHub is the world's largest collection of open source software, with around 28 million users and 79 million repositories.\footnote{\url{https://github.com/search}, verified on 03/05/2018.} In addition to a  {\texttt{git}-based version control system, GitHub integrates several features for social coding. For example, developers can \emph{fork} their own copy of a repository, work and improve the code locally, and then submit a \emph{pull request} to integrate the changes in the main repository~\cite{gousios2014exploratory, Gousios2015, Yu2015, GSB16}. 
%The key characteristics and challenges of this pull-based development model are explored in many studies
%Moreover, 
%GitHub also supports features typical of modern social networks. For example, 
Inspired by the {\em like} button of modern social networks, GitHub users can also \emph{star} a repository, presumably to manifest interest or satisfaction with the hosted project~\cite{begel2013}. However, the real and practical meaning of ``starring a project''  was never the subject of an in-depth and well-founded empirical investigation.

Furthermore, GitHub's success contributed to the emergence of a competitive open source market.
As a result, it is common to see projects competing for the same users. For example, {\sc AngularJS}, {\sc React}, and {\sc Vue.js} compete for developers of JavaScript single-page Web applications. This fact increases the relevance of studying the characteristics and practical value of  GitHub popularity metrics.\\

\begin{figure}[!t]
\centering
\includegraphics[width=0.7\linewidth, trim={0 1em 0 0}, clip]{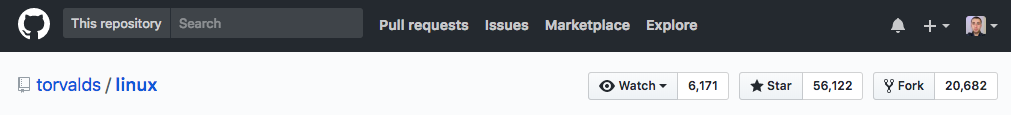}
\caption{GitHub popularity metrics}
\label{fig:intro:gh}
\end{figure}

\noindent{\em Motivating Survey:} In order to provide initial evidence on the most useful metrics for measuring the popularity of GitHub projects, we conducted a survey with Stack~Overflow users. We rely on these participants because Stack~Overflow is a widely popular programming forum, listing questions and answers about a variety of technologies, which are provided by practitioners with different profiles and background~\cite{vasilescu2013}. We randomly selected a sample of 400 Stack~Overflow users, using a dump of the site available on Google BigQuery.\footnote{\url{https://cloud.google.com/bigquery/public-data/stackoverflow}} We \mbox{e-mailed} these users asking then a single question: {\em How useful are the following metrics to assess the popularity of GitHub projects?} We then presented three common metrics provided by GitHub, which are displayed at the front page of any project: watchers, stars, and forks (see screenshot in Figure~\ref{fig:intro:gh}). Although available on any repository, project owners do not have control over  these metrics; any GitHub user can watch, star, or fork a repository, without asking permission to its owners. The survey participants were asked to rank the usefulness of these metrics in a 4-point Likert scale; we also configured the survey system to present the metrics in a random order, to avoid a possible order effect bias. We received 54 answers, which corresponds to a response ratio of 13.5\%. 

As presented in Figure~\ref{fig:intro:survey}, the results show that stars are viewed by practitioners as the most useful measure of popularity on GitHub, with 83\% of answers with scores 3 (31\%) or 4 (52\%). It is followed by forks with 72\% of answers with scores 3-4 (35\% and 37\%, respectively) and by watchers with 67\% (37\% and 30\%, respectively). Therefore, this initial survey confirms the importance of GitHub stars to practitioners, when compared to forks and watchers.
Additionally, stars are often used by researchers to select GitHub projects for empirical studies in software engineering~\cite{ray2014, padhye2014study, hilton2016usage, MazinanianKTD17, JIANG201744, Nielebock2018,Rigger2018, castro2018}. Therefore, a throughout analysis of starring practices can  shed light on the properties and risks involved in this selection.

\begin{figure}[!ht]
\centering
\includegraphics[width=0.6\linewidth, trim={0 1em 0 0}, clip]{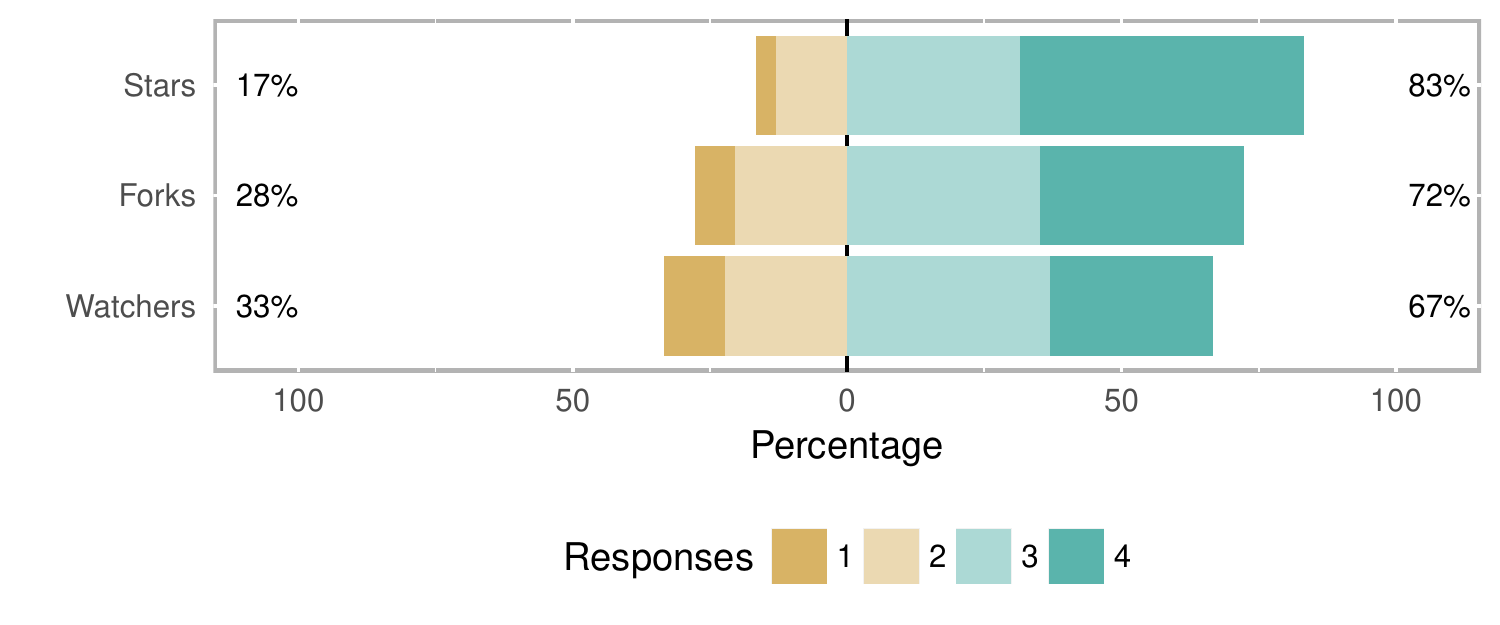}
\caption{How useful are the following metrics to assess the popularity of GitHub projects? (1: not useful; 4: very useful)}
\label{fig:intro:survey}
\end{figure}

\vspace{1em}\noindent{\em Proposed Study:} In a previous conference paper, we started an investigation on the factors and patterns that govern the number of stars of 2,500 GitHub projects~\cite{borges2016}.
Our first quantitative results indicate that: (i) repositories owned by organizations have more stars than the ones owned by individuals; (ii) there is no correlation between stars and repository's age; but there is a correlation with forks; (iii) repositories tend to receive more stars right after their public release; after this initial period, the growth rate tends to stabilize; (iv) there is an acceleration in the number of stars gained after releases.
Furthermore,  we showed that the growth of the number of stars is explained by four patterns, which we called {\em slow}, }{\em moderate}, {\em fast}, and {\em viral}.

In this paper, we extend this first study in three major directions:

\begin{enumerate}

\item We increment the number of systems from 2,500 to 5,000 public GitHub repositories.

\item We conduct two surveys to shed light on the quantitative results of the initial study. First, we perform a survey with 791 developers to reveal their motivations for starring projects. Ultimately, our intention is to understand {\em why} developers star GitHub projects. We also conduct a second survey with 115 project owners to reveal their perceptions about the growth patterns proposed in the first study. 

\item We investigate  the endogenous factors (i.e.,~the ones that can be extracted directly from a repository, like age) that affect the classification of a project in a given growth pattern. We collect 31 factors along three dimensions and use a machine learning classifier to identify the factors that most distinguish the projects across the proposed growth patterns.

\end{enumerate}

\noindent{\em Contributions:} Our work leads to at least five contributions:

\begin{enumerate}

\item To our knowledge, we are the first to provide solid empirical evidence---both quantitative and qualitative---on the meaning of the number of GitHub stars. Consequently, we recommend that open source maintainers should monitor this metric, as they monitor other project metrics, such as the number of pending issues or pull requests.

\item We reveal that active promotion, particularly on social media sites, has a key importance to increase the number of stars of open source projects. Since these projects are usually maintained by one or two contributors~\cite{icpc2016}, they should allocate time not only to write and maintain the code (developers role) but also to promote the projects (marketing role). 

%We also show the importance of using organizational accounts (e.g., {\em aserg-ufmg} instead of {\em hsborges}).

\item We distill a list of threats faced by practitioners and researchers when selecting GitHub projects based on the number of stars. For example, this selection may favor projects with active marketing and advertising strategies, instead of projects following well-established software engineering practices. 

\item We implement an open source tool (\url{http://gittrends.io}) to explore and check our results, including the time series of stars used in this paper and the proposed growth patterns. 

\item We provide a public dataset (\url{https://doi.org/10.5281/zenodo.1183752}) with the application domain of 5,000 GitHub repositories. This dataset can support research in a variety of Software Engineering problems and contexts.

\end{enumerate}

\noindent{\em Structure:} 
Section~\ref{sec:dataset} presents and characterizes the dataset used in the study.
Section~\ref{sec:survey} reports GitHub users' major motivations for starring repositories.
Section~\ref{sec:characterization} presents a quantitative study on the number of stars of GitHub repositories.
Section~\ref{sec:patterns} documents the patterns we propose to describe the growth of the number of stars of GitHub systems.
Section~\ref{sec:factors} investigates factors potentially affecting the inclusion of a repository in the proposed growth patterns.
Section~\ref{sec:patterns-survey} describes project owners' perceptions about the growth patterns of their repositories.
Threats to validity are discussed in Section~\ref{sec:threats} and related work is presented in Section~\ref{sec:related-work}.
We conclude by summarizing our findings and listing future work in Section~\ref{sec:conclusion}.

% !TEX root = jss.tex
\section{Dataset}
\label{sec:dataset}

The dataset used in this paper includes the top-5,000 public repositories by number of stars on GitHub.
We limit the study to 5,000 repositories for two major reasons. First, to focus on the characteristics of the most starred GitHub projects.
Second, because we investigate the impact of application domain on number of stars, which demands a manual classification of each system domain.

All data was obtained using the GitHub API, which provides services to search public repositories and to retrieve specific data about them (e.g., stars, commits, contributors, and forks). The data was collected on January 23rd, 2017.
Besides retrieving the number of stars for each system, we also relied on the GitHub API to collect historical data about the number of stars. For this purpose, we used a service from the API that returns all events of a given repository. For each star, these events store the date and the user who starred the repository. However, the GitHub API returns at most 100 events by request (i.e., a page) and at most 400 pages. For this reason, it is not possible to retrieve all stars events of systems with more than 40K stars, as is the case for 18 repositories, such as \textsc{FreeCodeCamp}, \textsc{Bootstrap}, \textsc{D3}, and \textsc{Font-Awesome}. Therefore, these 18 systems are not considered in Sections~\ref{sec:patterns},~\ref{sec:factors}, and~\ref{sec:patterns-survey}, since we depend on the complete time series to cluster and derive stars growth patterns.

Table~\ref{tab:top10-repositories-statistics} shows descriptive statistics on the number of stars of the repositories in our dataset. The number of stars ranges from 1,596 (for \textsc{mapnik/mapnik}) to 224,136 stars (for \textsc{FreeCodeCamp/FreeCodeCamp}). The median number of stars is 2,866.

\begin{table}[!ht]
\centering
\caption{Descriptive statistics on the number of stars of the repositories in our dataset}
\label{tab:top10-repositories-statistics}
\begin{tabular}{@{}ccccc@{}}
\toprule
Min & 1st Quartile & 2nd Quartile & 3rd Quartile & Max \\
\midrule
1,596 & 2,085 & 2,866 & 4,541 & 224,136 \\
\bottomrule
\end{tabular}
\end{table}

\begin{figure}[!ht]
\centering
\begin{subfigure}[t]{0.225\textwidth}
\includegraphics[width=\textwidth,trim={0 2em 0 0},clip]{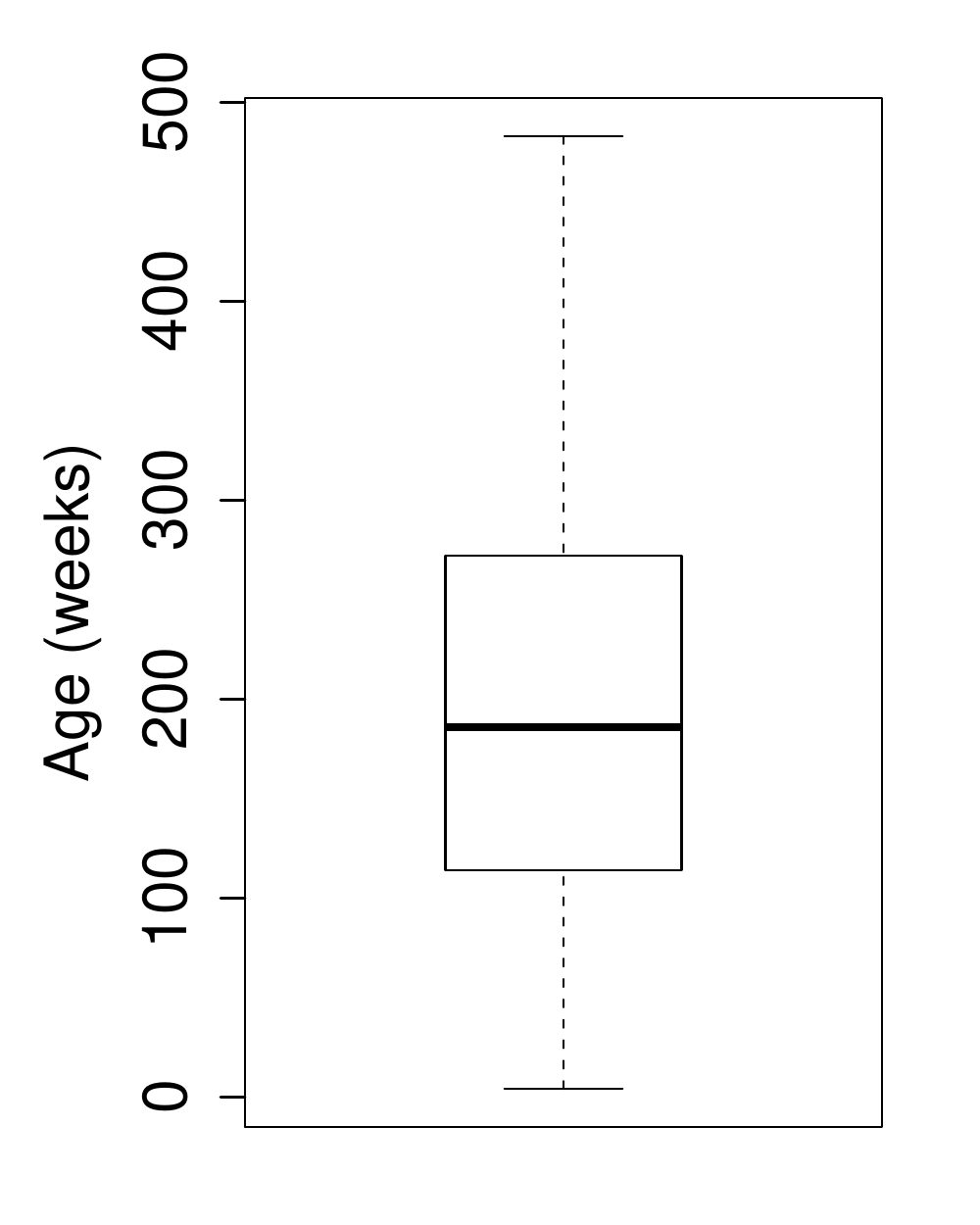}
\caption{Age (weeks)}
\label{fig:age-overview}
\end{subfigure}%
\begin{subfigure}[t]{0.225\textwidth}
\includegraphics[width=\textwidth,trim={0 2em 0 0},clip]{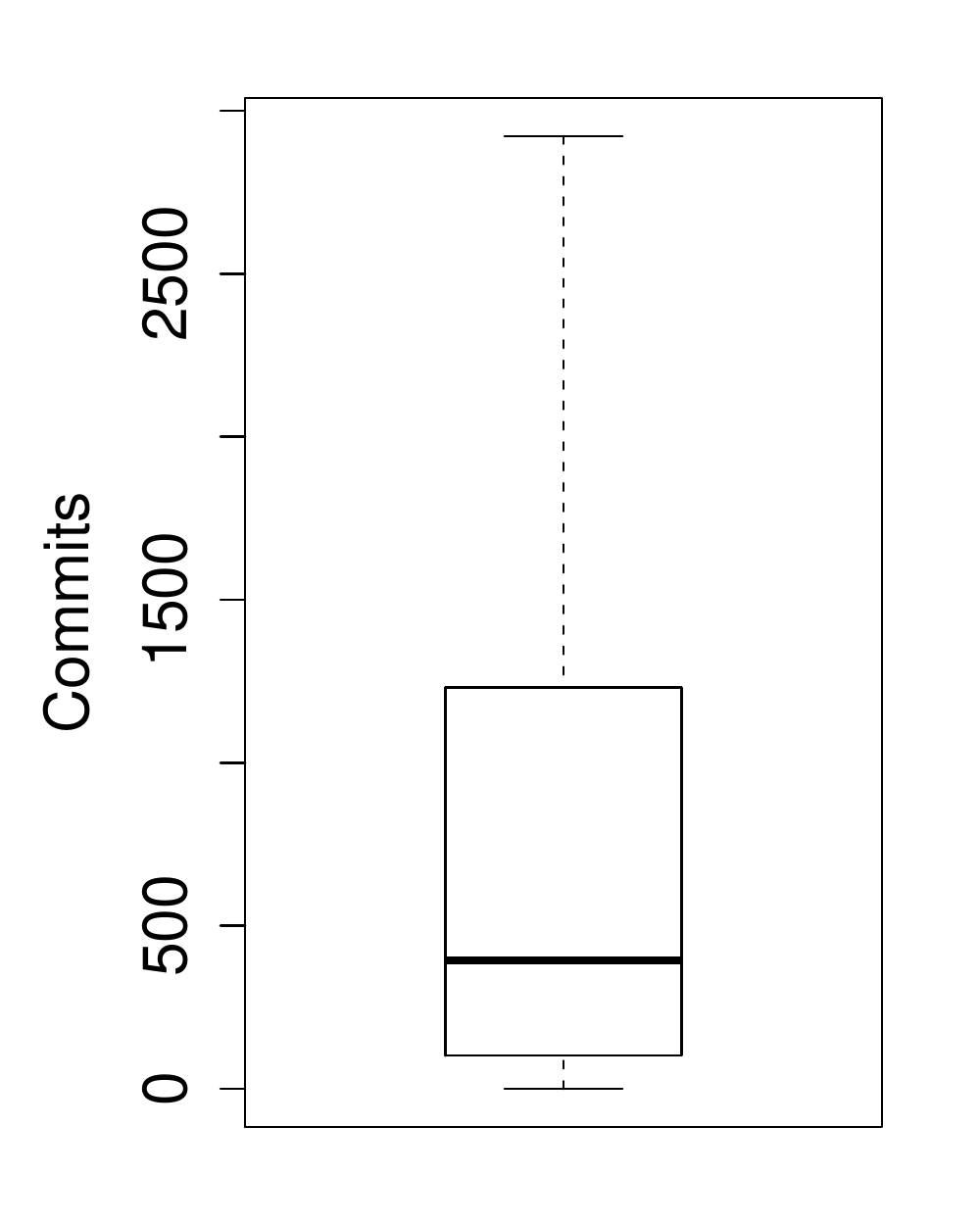}
\caption{Commits}
\label{fig:commits-overview}
\end{subfigure}
\begin{subfigure}[t]{0.225\textwidth}
\includegraphics[width=\textwidth,trim={0 2em 0 0},clip]{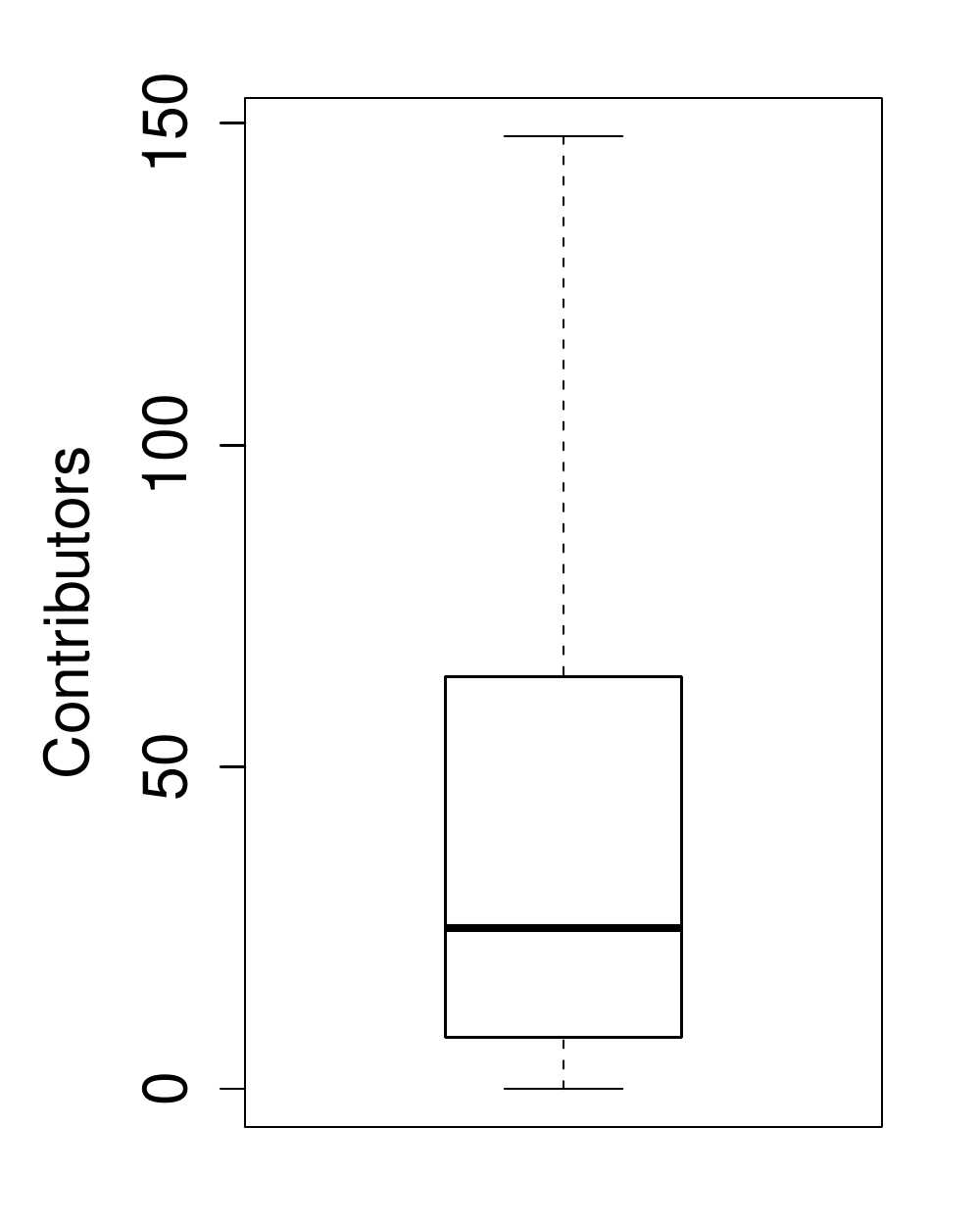}
\caption{Contributors}
\label{fig:size-overview}
\end{subfigure}%
\begin{subfigure}[t]{0.225\textwidth}
\includegraphics[width=\textwidth,trim={0 2em 0 0},clip]{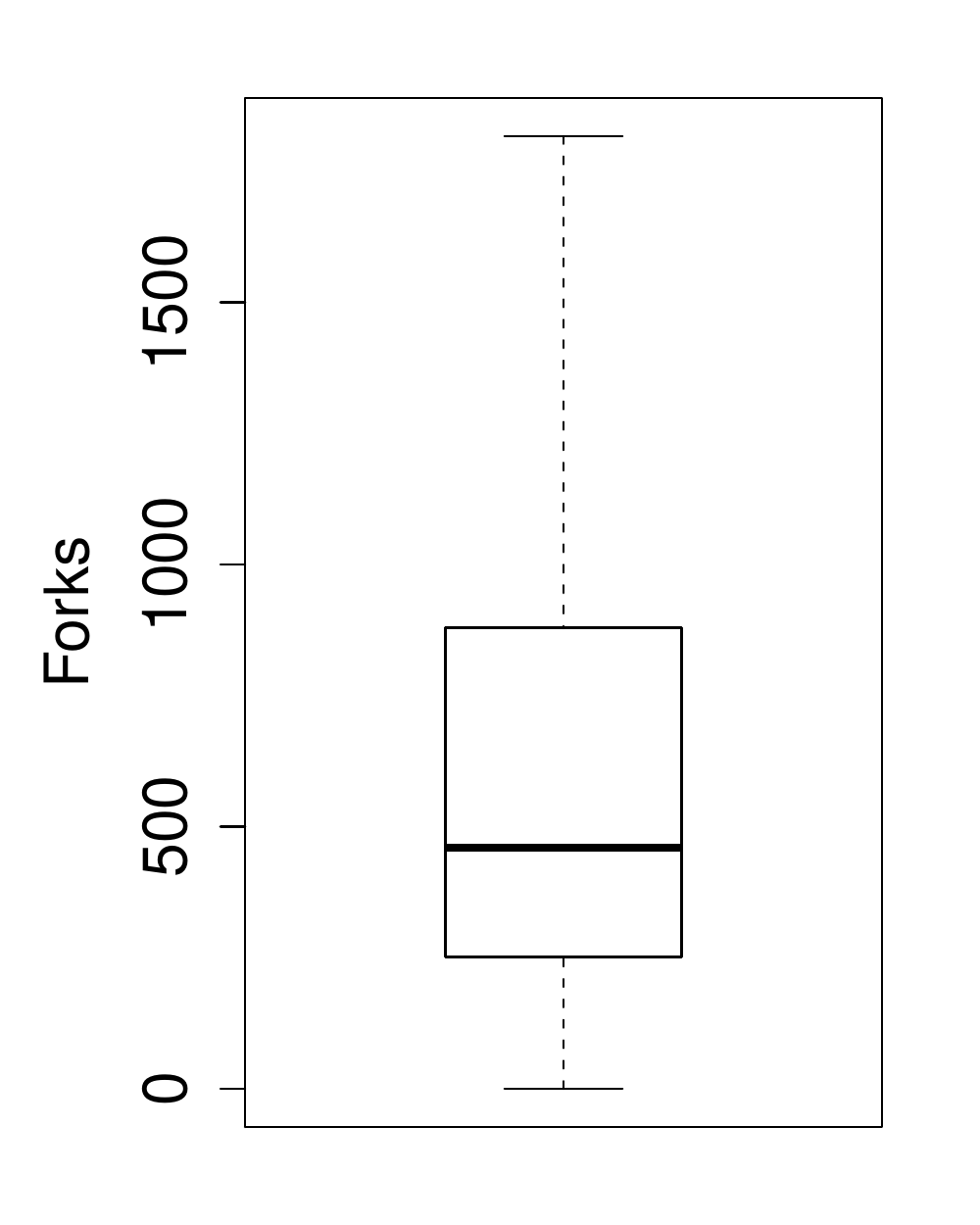}
\caption{Forks}
\label{fig:forks-overview}
\end{subfigure}
\caption{Age, number of commits, number of contributors, and number of forks (outliers are omitted)}
\label{fig:dataset-info}
\end{figure}

\vspace{1em}\noindent\emph{Age, Commits, Contributors, and Forks:} Figure~\ref{fig:dataset-info} shows boxplots with the distribution of the age (in number of weeks), number of commits, number of contributors, and number of forks for the 5,000 systems in the dataset.
For age, the first, second, and third quartiles are 114, 186, and 272 weeks, respectively.
For number of commits, the first, second, and third quartiles are 102, 393, and 1,230, respectively.
For number of contributors, the first, second, and third quartiles are 8, 25, and 64, respectively;\footnote{We report contributors data as retrieved by the GitHub API. This data may be different from the one presented on the project's page on GitHub, which only counts contributors with GitHub account.} and for number of forks, the first, second, and third quartiles are 252, 460, and 879, respectively. Therefore, the systems in our dataset usually have years of development and many commits and contributors.

\begin{figure}[!ht]
\centering
\includegraphics[width=0.7\linewidth, trim={0 1em 0 0}, clip]{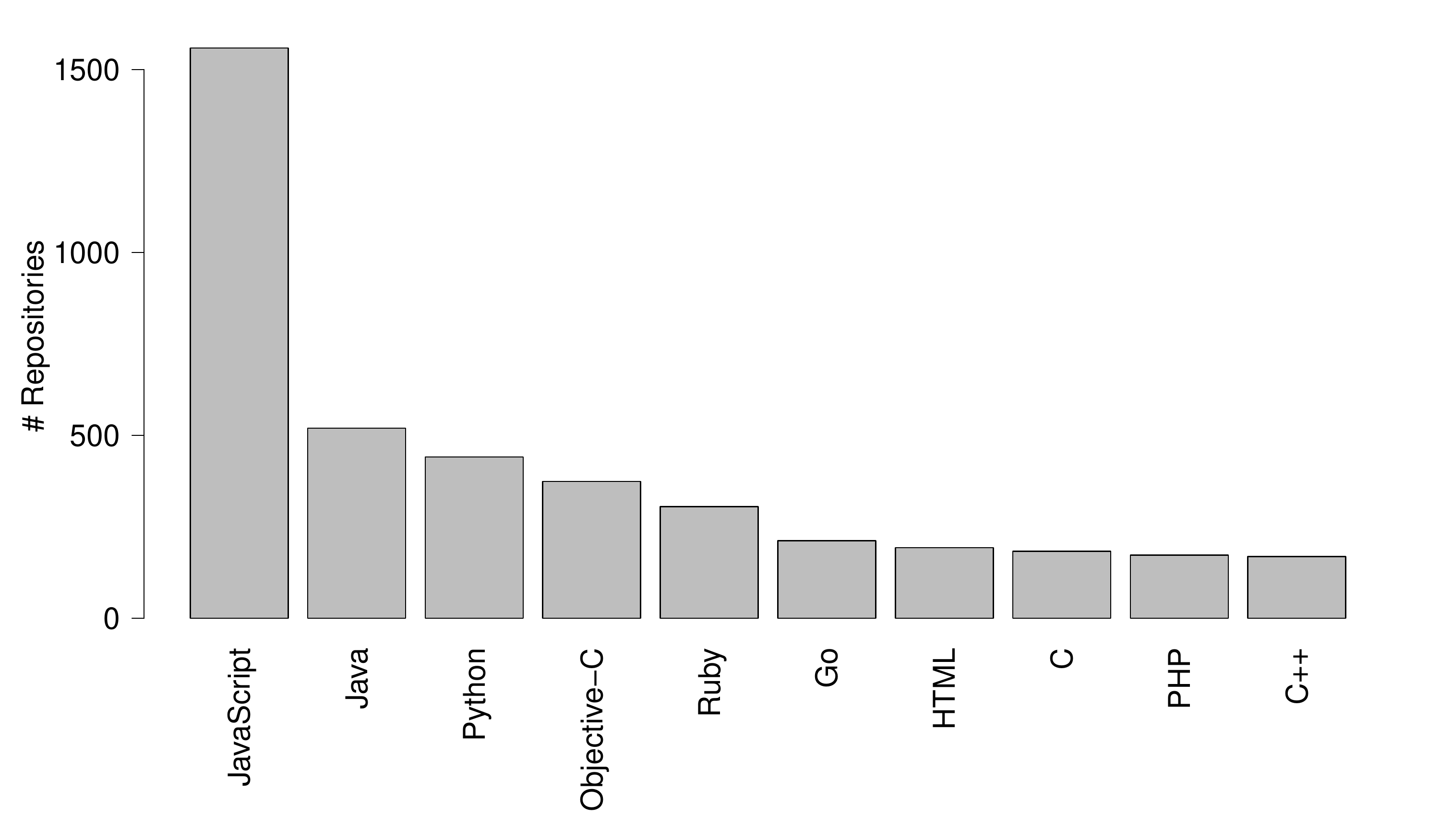}
\caption{Top-10 languages by number of repositories}
\label{fig:stars-overview2}
\end{figure}

\vspace{1em}\noindent\emph{Programming Language:} As returned by the GitHub API, the language of a project is the one with the highest percentage of source code in its repository. Figure~\ref{fig:stars-overview2} shows the distribution of the systems per programming language. JavaScript is the most popular language (1,559 repositories, 31.1\%), followed by Java (520 repositories, 10.4\%), Python (441 repositories, 8.8\%), Objective-C (374 repositories, 7.4\%), and Ruby (305 repositories, 6.1\%). Despite a concentration of systems in these languages, the dataset includes systems in 71 languages, including Cuda, Julia, SQLPL, and XSLT (all with just one repository).\footnote{Although HTML is a markup language, it is included in Figure~\ref{fig:stars-overview2}. The reason is that we also intend to study repositories containing documentation.}

\vspace{1em}\noindent\emph{Owner:} We also characterize our dataset according to repository owner. On GitHub, a repository can be owned by a user (e.g.,~\textsc{torvalds/linux}) or by an organization (e.g.,~\textsc{facebook/react}). In our dataset, 2,569 repositories (51.3\%) are owed by users and 2,431 repositories (48.7\%) by organizations.

\vspace{1em}\noindent\emph{Application Domain:} In this study, we also group repositories by application domain. However, different from other source code repositories, like SourceForge, GitHub does not include information about the application domain of a project. For this reason, we manually classified the domain of each system in our dataset. Initially, the first author of this paper inspected the description of the top-200 repositories to provide a first list of application domains, distributed over six domain types, as presented next. These domains were validate with the second paper's author. After this initial classification, the first author inspected the short description, the GitHub page and the project's page of the remaining 4,800 repositories. During this process, he also marked the repositories with dubious classification decisions. These particular cases were discussed by the first and second authors, to reach a consensus decision. To the best of our knowledge, this is the first large-scale classification of application domains on GitHub.

The systems are classified in the following six domains:\footnote{This classification only includes first-level domains; therefore, it can be further refined to include subdomains, such Android vs desktop applications.}

\begin{enumerate}
\item Application software: systems that provide functionalities to end-users, like browsers and text editors (e.g.,~\textsc{WordPress/WordPress} and \textsc{adobe/brackets}).
\item System software: systems that provide services and infrastructure to other systems, like operating systems, middleware, and databases (e.g.,~\textsc{torvalds/linux} and \textsc{mongodb/mongo}).
\item Web libraries and frameworks: systems that are used to implement the front-end (interface) of web-based applications (e.g.,~\textsc{twbs/bootstrap} and \textsc{angular/angular.js}).
\item Non-web libraries and frameworks: systems that are used to implement other components of an application, despite a web-based interface (e.g.,~\textsc{google/guava} and \textsc{facebook/fresco}).
\item Software tools: systems that support development tasks, like IDEs, package managers, and compilers (e.g.,~\textsc{Homebrew/homebrew} and \textsc{git/git}).
\item Documentation: repositories with documentation, tutorials, source code examples, etc. (e.g.,~\textsc{iluwatar/\\java-design-patterns}).
\end{enumerate}

\begin{figure}[!ht]
\centering
\includegraphics[width=.5\linewidth, trim={0 0 0 0}, clip]{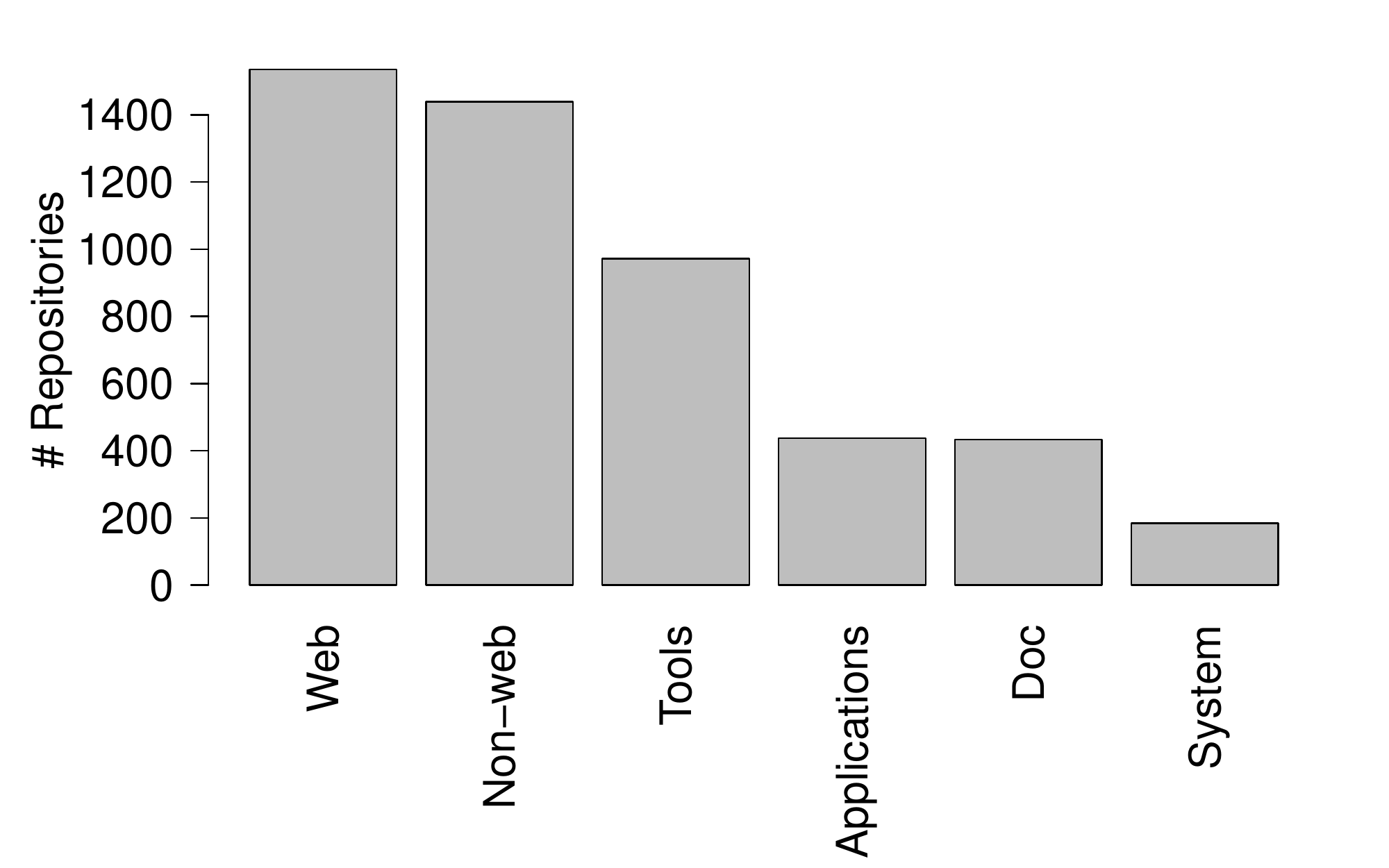}
\caption{Number of repositories by domain}
\label{fig:domains-dist}
\end{figure}

Figure~\ref{fig:domains-dist} shows the number of systems in each domain. The top-3 domains are web libraries and frameworks (1,535 repositories, 30.7\%), non-web libraries and frameworks (1,439 repositories, 28.7\%), and software tools (972 repositories, 19.4\%). The projects in these domains can be seen as meta-projects, i.e.,~they are used to implement other projects, in the form of libraries, frameworks, or documentation.

% !TEX root = jss.tex
% !TeX spellcheck = en-US
\section{Survey Study}
\label{sec:survey}

In this section, we describe an investigation with developers to reveal their motivations for starring projects and to check whether they consider the number of stars before using or contributing to projects on GitHub.
Section~\ref{sec:survey:design} describes the design of the survey questionnaire and the selection of the survey participants.
Section~\ref{sec:survey:results} reports the survey results.

\subsection{Survey Design}
\label{sec:survey:design}

The survey questionnaire has two open-ended questions: (1) Why did you star \textit{owner/name}? and (2) Do you consider the number of stars before using or contributing to a GitHub project? In the first question, \textit{owner/name} refers to a repository. Our intention with this question is to understand the motivations behind a developer's decision to star a GitHub repository.
%investigate whether stars can be viewed as a measure of popularity, which we define as follows: ``the state of being liked, enjoyed, accepted, or done by a large number of people'', according to Merriam-Webster. 
With the second question, our goal is to check whether stars is indeed a factor considered by developers when establishing a more close relationship with a project, as a client (or user) or as a contributor (or developer).
These questions were sent by email to the last developer who starred each repository in our dataset.
The emails were obtained using the GitHub API. When the developers who gave the last star do not have a public email, we select the previous one and so on, successively.
We excluded 276 repositories (5.5\%) because the last star was given more than six months before the data collection.
Therefore, this increases the probability of developers not remembering the concrete reasons they starred these repositories.
Moreover, for 336 repositories (6.7\%), the selected developer also recently starred other repository in our dataset, thus we excluded these repositories to avoid sending multiple emails to the same developer.
Finally, our sample of participants consists of 4,370 developers who recently starred 4,370 repositories from our dataset.

The questionnaire was sent between 13rd and 27th of March 2017.
After a period of 30 days, we obtained 791 responses and 173 e-mails returned due to delivery issues (e.g., non-existent recipient), resulting in a response rate of 18.8\%. This number of answers represent  a confidence interval of 3.15\%, 
for a confidence level of 95\%.
Considering the locations configured in the respondents' GitHub profile, 133 respondents (16.8\%) are from the United States, 74 respondents (9.4\%) are from China, 39 (4.9\%) are from Brazil, 34 (4.3\%) are from Canada, and 27 (3.4\%) from India. Other 321 respondents (40.6\%) are from 68  different countries and 163 respondents (20.6\%) have no location configured in their GitHub profiles.
Regarding the respondents' experience in the GitHub platform, their account age ranges from 18 days to 9.12 years, with an average of 4.16 years and a median of 4.09 years. Regarding the programming language used by the participants, 32.6\%  have most of their public GitHub projects implemented in JavaScript, followed by Python (12.5\%), Java (11.8\%), Ruby (7.0\%), and PHP (5.0\%).

To preserve the respondents privacy, we use labels P1 to P791 when quoting the answers.
We analyze the answers using thematic analysis~\cite{cruzes2011}, a technique for identifying and recording \aspas{themes} (i.e., patterns) in textual documents. Thematic analysis involves the following steps: (1) initial reading of the answers, (2) generating a first code for each answer, (3) searching for themes among the proposed codes, (4) reviewing the themes to find opportunities for merging, and (5) defining and naming the final themes. All steps were performed by the first author of this paper.

\subsection{Survey Results}
\label{sec:survey:results}

This section presents the answers to the survey questions. A separate subsection discusses each question.

\subsubsection{Why did you star \textit{owner/name}?}

In this question, we asked the developers to respond why they starred a given repository.
In the next paragraphs, we present four major reasons that emerged after analysing the answers.\bigskip

\noindent{\textbf{To show appreciation:}} More than half of the participants (52.5\%) answered they starred the repositories because they liked the project.
In general, the answers mention that stars are used as \aspas{likes} button in other social networks, such as Facebook and YouTube.
As examples we have:\medskip

\noindent{\textit{I liked the solution given by this repo.}} (P373)\smallskip

\noindent{\textit{I starred this repository because it looks nice.}} (P689)\bigskip

\noindent{\textbf{Bookmarking:}} 51.1\% of the participants reported they starred the repositories for later retrieval.
We have the following answers as examples:\medskip

\noindent{\textit{I starred it because I wanted to try using it later.}} (P250)\smallskip

\noindent{\textit{Because I use stars as a \aspas{sort of} bookmarks.}} (P465)\bigskip

\noindent{\textbf{Due to usage:}} 36.7\% of the participants reported they used or are using the project.
As examples we have:\medskip

\noindent{\textit{I have been using for many years and was about to use again in a new project.}} (P162)\smallskip

\noindent{\textit{Because it solved my problem.}} (P650)\bigskip

\noindent{\textbf{Due to third-party recommendations:}} 4.6\% of the participants starred the repositories due to recommendations from friends, websites, or other developers, as in this answer:\medskip

\noindent{\textit{I starred the repository because a technological group recommended it.}} (P764)\bigskip

Additionally, five developers (0.6\%) answered they do not know or remember the reason why they starred the repositories.
Table~\ref{tab:users:q1} details the number of answers and the percentage of responses on each theme.
Note that one answer can receive more than one theme.
For example, the theme \textit{To show appreciation} appeared together with \textit{Bookmarking} and \textit{Due to usage} in 122 and 116 answers, respectively.
Moreover, \textit{Due to usage} and \textit{Bookmarking} appeared together in 63 answers.

\begin{table}[!ht]
\centering
\caption{Why do users star GitHub repositories?\\(95\% confidence level with a 3.15\% confidence interval)}
\label{tab:users:q1}
\begin{tabular}{@{}lrc@{}}
\toprule
\multicolumn{1}{c}{Reason} & \multicolumn{1}{c}{Total} & \multicolumn{1}{c}{\%} \\
\midrule
To show appreciation   & 415 & 52.5 \sbarf{415}{791} \\
Bookmarking            & 404 & 51.1 \sbarf{404}{791} \\
Due to usage           & 290 & 36.7 \sbarf{290}{791} \\
Due to recommendations &  36 & \enspace4.6 \sbarf{36}{791}  \\
Unknown reasons        &   5 & \enspace0.6 \sbarf{5}{791}   \\
\bottomrule
\end{tabular}
\end{table}

\vspace{1em}\begin{tcolorbox}[left=0.75em,right=0.75em,top=0.75em,bottom=0.75em,boxrule=0.25mm,colback=gray!5!white]
\noindent{\em Summary:} GitHub developers star repositories mainly to show appreciation to the projects (52.5\%), to bookmark projects for later retrieval (51.1\%), and because they used or are using the projects (36.7\%).
\end{tcolorbox}\vspace{1em}

\subsubsection{Do you consider the number of stars before using or contributing to a project?}
\label{sec:survey:results:q2}

In the second question, we asked the participants to respond if they consider the number of stars before using or contributing to GitHub projects.\footnote{Therefore, in this survey, we do not distinguish usage and contribution to Github repositories, which is left for future work.}
From the 791 answers received in the survey, 14 developers (1.7\%) did not answer this specific question. Thus, the numbers presented in this section refer to 777 responses, which gives an updated confidence interval of 3.19\%, for a confidence level of 95\%.
First, we classified the answers in \textit{yes} (the participant does consider the number of stars) and \textit{no} (the participant does not consider the number of stars).
As observed in Table~\ref{tab:users:q2a}, 73\% of the participants consider the number of stars before using or contributing to GitHub projects and 23.3\% answered negatively to this question.
Finally, 3.7\% of the participants did not correctly answer the question, probably due to a misunderstanding. For example, participant P745 just provided the following answer: \aspas{\textit{I am not an active OSS contributor}}.

\begin{table}[!ht]
\centering
\caption{Do GitHub users consider the number of stars before using or contributing to a project? (95\% confidence level with\\ a 3.19\% confidence interval)}
\label{tab:users:q2a}
\begin{tabular}{@{}crc@{}}
\toprule
\multicolumn{1}{c}{Answer} & \multicolumn{1}{c}{Total} & \multicolumn{1}{c}{\%} \\
\midrule
Yes & 567 & 73.0 \sbarf{730}{1000} \\
No & 181 & 23.3 \sbarf{233}{1000} \\
Unclear & 29 & \enspace3.7 \sbarf{37}{1000} \\
\bottomrule
\end{tabular}
\end{table}

\noindent\textbf{Positive Answers:} Considering the participants who answered positively to this second question, 26.5\% commented that the number of stars has a high influence on their decision of using or contributing to a project.
As examples, we have these answers:\medskip

\noindent{\textit{I always consider the amount of stars on a repo before adopting it in a project. It is one of the most important factors, and in my opinion gives the best metric at a glance for whether a package is production ready.}} (P365)\medskip

\noindent{\textit{Of course stars count is very useful thing, because it tells about project quality. If many people starred something - many people think that it is useful or interesting.}} (P31)\bigskip

For 29.3\% of the participants who provided a positive answer, the number of stars is just one of the factors they consider before using or contributing to GitHub projects. Other factors include quality of the code/documentation, recent activity, license, and project owner.
As examples, we have the following answers:\medskip

\noindent{\textit{Yes. I do not take it as my only metric, but having a considerable number of stars and recent activity is reassuring in terms of it being a stable project that my projects can depend on in future.}} (P104)\medskip

\noindent{\textit{I often consider the number of stars (as well as recency of commits, PRs, and issues) in deciding whether to use a project.}} (P442)\bigskip

Moreover, 8.8\% of the participants consider the number of stars when using but not when contributing to GitHub projects. For example:\medskip

\noindent{\textit{I usually contribute more to projects with less stars because of the ease of approach to a smaller community, hence project. On the other hand I normally use frameworks with more stars because of the continuous support they have.}} (P642)\bigskip

Additionally, 46 participants (8.1\%) provided other comments, as in the following answers:\medskip

\noindent{\textit{Yes, a little, I look if it has at least a couple of stars to be sure that doesn't get unmaintained in a short term}} (P89)\medskip

\noindent{\textit{Number of stars is not the major point for me. But it can serve as indicator of something really good}} (P224)\medskip

\noindent{\textit{I don't really notice exactly how many stars something has, but I do notice orders of magnitude (hundreds vs thousands vs tens of thousands)}} (P421)\bigskip

Finally, 194 developers (34.2\%) did not provide additional information to justify their positive answers.\bigskip

\noindent\textbf{Negative Answers:} Considering only the participants who answered negatively to this second question, 45 participants (24.9\%) commented they consider the purpose, domain, and features of the project, but not the number of stars. As examples, we have the answers:\medskip

\noindent{\textit{No, my primary interest is: what problem is solving by this project}} (P203)\medskip

\noindent{\textit{Not really. If I like the strategy and implementation, I don't really care how popular or unpopular the repository is}} (P560)\bigskip

Moreover, 38 developers (21.0\%) answered they consider other measures and sources of information on their decisions, but not the number of stars. For example:\medskip

\noindent{\textit{No, I don't consider the number of stars. Number of contributors, commits are important instead of number of stars}} (P270)\medskip

\noindent{\textit{No, I usually know a project from a different source than GitHub itself so I rather refer to the outside opinions on a framework (blogs, articles, community, \ldots) on whether it is of good quality than a stars on GitHub}} (P557)\bigskip

Additionally, 26 participants (14.3\%) provided other reasons for not considering the number of stars (e.g., stars do not reflect project quality); and 74 developers (40.8\%) did not provide additional information to justify their answers.

\vspace{1em}\begin{tcolorbox}[left=0.75em,right=0.75em,top=0.75em,bottom=0.75em,boxrule=0.25mm,colback=gray!5!white]
\noindent{\em Summary:} Three out of four developers consider the number of stars before using or contributing to GitHub projects. Among the developers who consider stars, 29.3\% also evaluate other factors, such as source code quality, license, and documentation.
\end{tcolorbox}\vspace{1em}

% !TEX root = jss.tex
\section{Characterization Study}
\label{sec:characterization}

In this section, we describe a quantitative characterization of the number of stars of GitHub projects.\footnote{This section and the next one are based in our previous conference paper~\cite{borges2016}, but increasing the number of analysed systems from 2,500 to 5,000 open source projects.}
More specifically, we provide answers to four research questions:\medskip

\noindent \emph{RQ \#1: How the number of stars varies per programming language, application domain, and repository owner?} The goal is to provide an initial view about the number of stars of the studied systems, by comparing this measure across programming language, application domain, and repository owner (user or organization).\medskip

\noindent \emph{RQ \#2: Does stars correlate with repository's age, number of commits, number of contributors, and number of forks?} This investigation can help to unveil possible selection bias that occurs when ranking projects based on the number of stars. For example, a positive correlation with repository's age would imply that ranking by stars favors older projects.
\medskip

\noindent \emph{RQ \#3: How early do repositories get their stars?} With this research question, we intend to check whether gains of stars are concentrated in specific phases of a repository's lifetime, specifically in early releases.\medskip

\noindent \emph{RQ \#4: What is the impact of new features on stars?} This investigation can show if relevant gains in the number of stars in the weeks following new releases.

%The proposed research questions aim to shed light on the relation between GitHub stars and other project metrics and characteristics.
%Some of our findings directly support actionable guidelines, e.g.,~we reveal that repositories owned by organizational accounts have more stars. Others motivate or support developers when making a decision to start a new open source project, e.g.,~we show that there is still a relevant demand for Web libraries and frameworks, mostly implemented in JavaScript. Furthermore, for existing repositories, our findings can explain their number of stars, but cannot be used to improve it; for example, it is not practical to migrate a project to a new programming language and, most obviously, to a new application domain.

\subsection{Results}
\label{sec:characterization:results}

\vspace{0.5em}\noindent\emph{RQ \#1: How the number of stars varies per programming language, application domain, and repository owner?}\vspace{0.5em}
\label{sub:results:rq2}

\begin{figure}[!ht]
\centering
\includegraphics[width=.6\linewidth, trim={0 0 0 0}, clip]{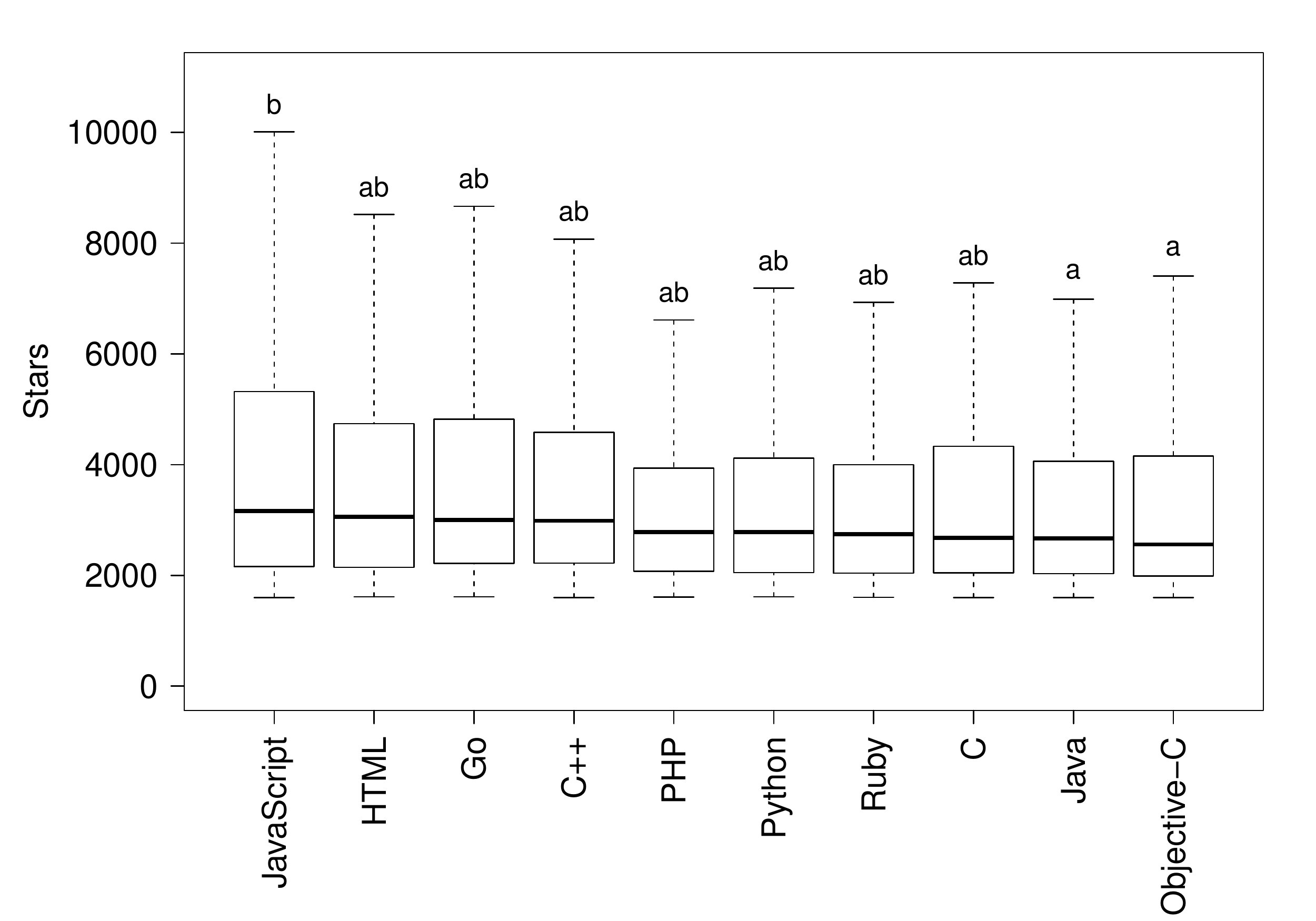}
\caption{Stars by programming language (considering only the top-10 languages with more repositories)}
\label{fig:language-popularity}
\end{figure}

Figure~\ref{fig:language-popularity} shows the distribution of the number of stars for the top-10 languages with more repositories.
The top-3 languages whose repositories have the highest median number of stars are: JavaScript (3,163 stars), HTML (3,059 stars), and Go (3,000 stars).
The three languages whose repositories have the lowest median number of stars are C (2,679 stars), Java (2,666 stars), and Objective-C (2,558 stars).
By applying the Kruskal-Wallis test to compare multiple samples, we found that these distributions differ in at least one language (\emph{p-value} $<$ 0.001).
Then, a non-parametric, pairwise, and multiple comparisons test (Dunn’s test)  was used to isolate the languages that differ from the others.
In Figure~\ref{fig:language-popularity}, the labels \textit{a} and \textit{b} in the bars express the results of Dunn's test.
Bars sharing the same labels indicate distributions that are not significantly different (\emph{p-value} $\leq$ 0.05).
For example, both JavaScript and HTML share the label \textit{b}, which means that these distributions have no statistical difference.
On the other hand, the distribution with the number of stars of JavaScript projects (label \textit{b}) is statistically different from Java (label \textit{a}).

\begin{figure}[!ht]
\centering
\includegraphics[width=.6\linewidth, trim={0 0 0 0}, clip]{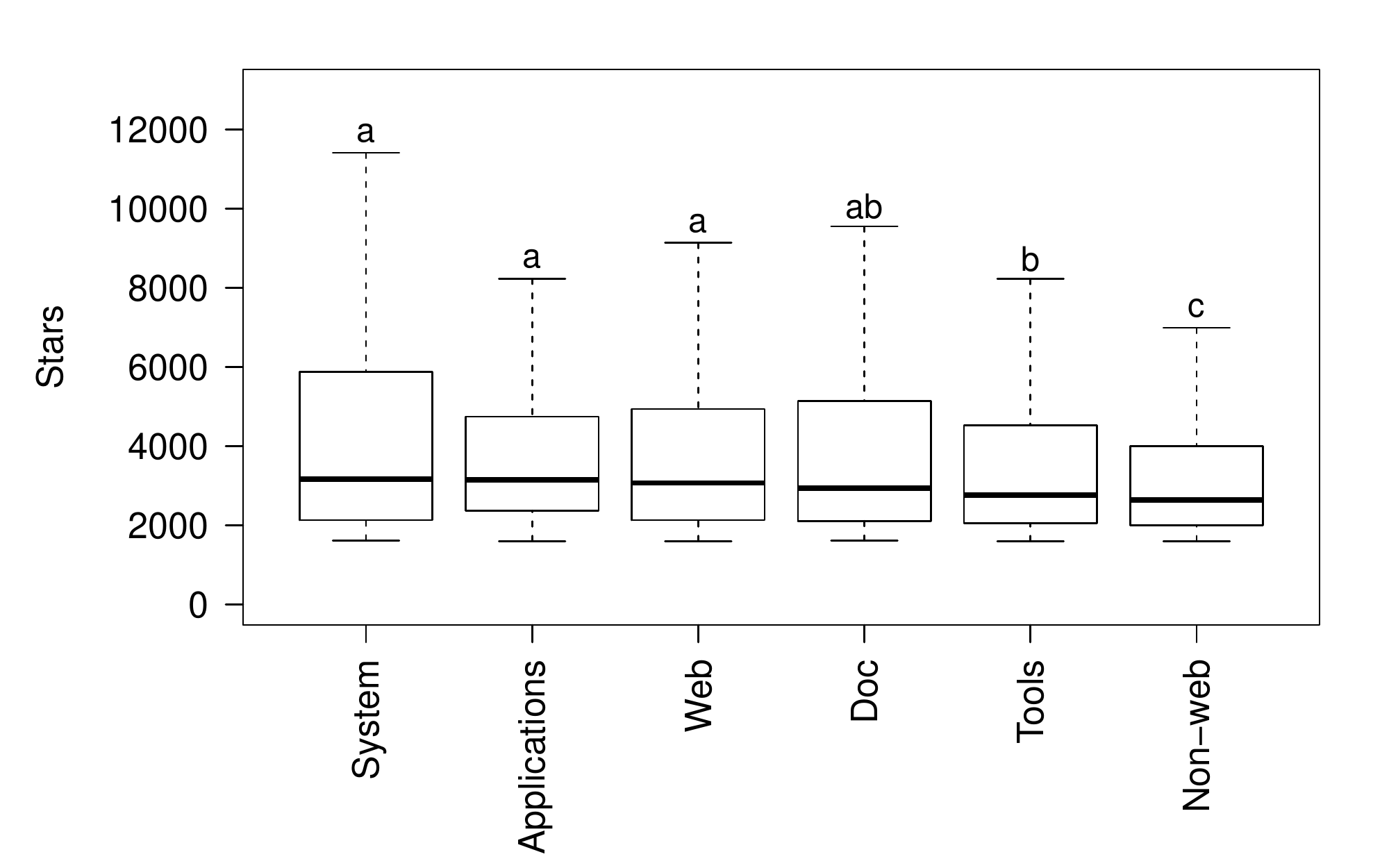}
\caption{Number of stars by application domain}
\label{fig:domains-popularity}
\end{figure}

Figure~\ref{fig:domains-popularity} shows the distribution of the number of stars for the repositories in each application domain.
The median number of stars varies as follow: systems software (3,168 stars), applications (3,147 stars), web libraries and frameworks (3,069 stars), documentation (2,942 stars), software tools (2,763 stars), and now-web libraries and frameworks (2,642 stars).
By applying the Kruskal-Wallis test, we found that the distributions are different (\emph{p-value} $<$ 0.001).
According to Dunn's test, the distribution of non-web libraries and frameworks (label \textit{c}) is statistically different from all other domains, showing that projects in this domain have less stars.
Similarly, tools (label \textit{b}) have more stars only than non-web libraries and frameworks (label \textit{c}).
Finally, there is no statistical difference between the number of stars of systems software, applications, web libraries and frameworks, and documentation (since all these distributions have the label \textit{a} in common).

\begin{figure}[!ht]
\centering
\includegraphics[width=.35\linewidth, trim={0 0 0 0}, clip]{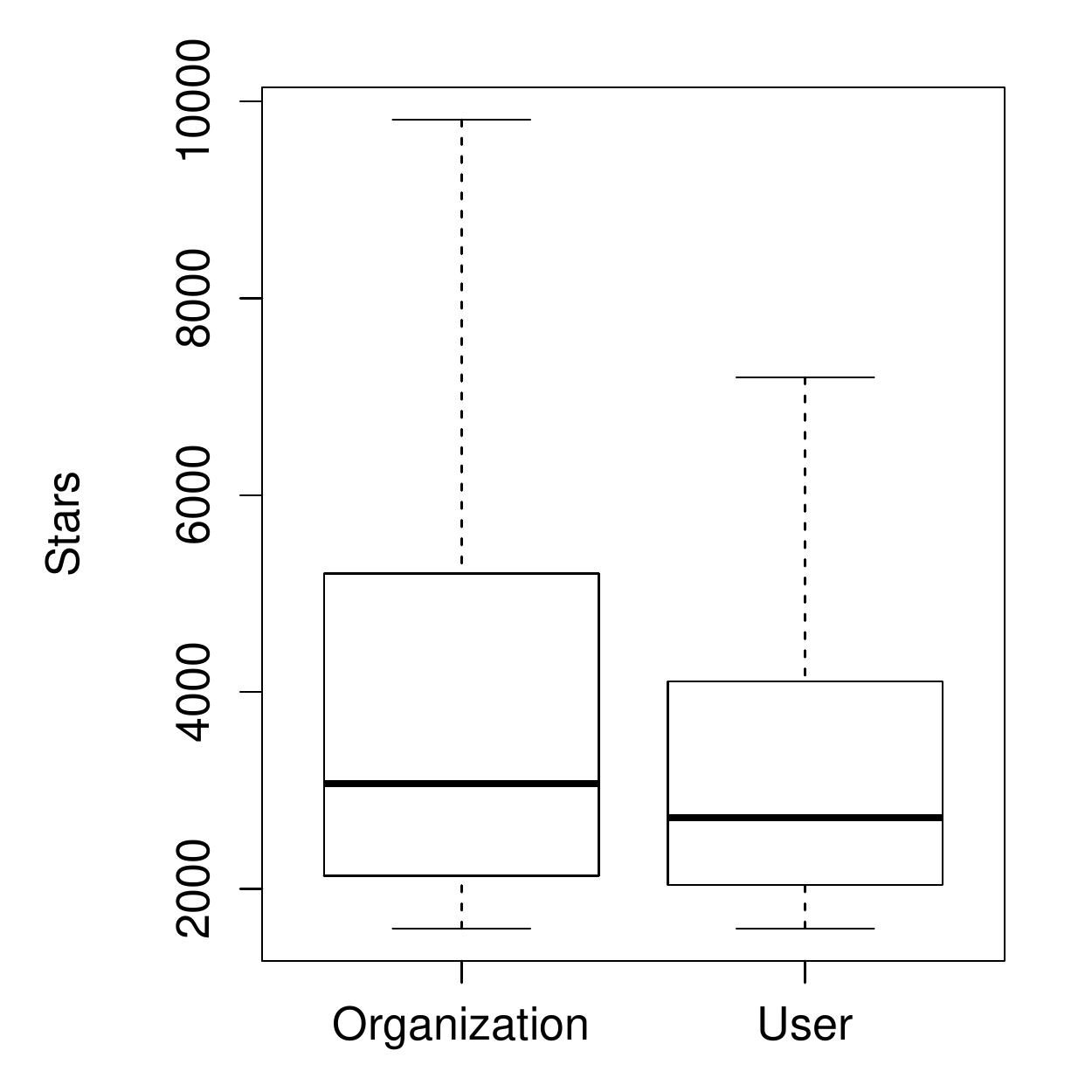}
\caption{Number of stars by repository owner}
\label{fig:owner-popularity}
\end{figure}

Finally, Figure~\ref{fig:owner-popularity} shows how the number of stars varies depending on the repository owner (i.e., user or organization).
The median number of stars is 3,067 stars for repositories owned by organizations and 2,723 stars for repositories owned by users.
By applying the Mann-Whitney test, we detected that these distributions are different (\emph{p-value} $<$ 0.001) with a {\em very small} effect size (Cohen's $d = -0.178$).
Our preliminary hypothesis is that repositories owned by organizations---specifically major software companies and free software foundations---have more funding and resources, which contributes to their higher number of stars.\bigskip

\vspace{0.5em}\begin{tcolorbox}[left=0.75em,right=0.75em,top=0.75em,bottom=0.75em,boxrule=0.25mm,colback=gray!5!white]
\noindent{\em Summary:} JavaScript repositories have the highest number of stars (median of 3,163 stars); and non-web libraries and frameworks have the lowest number (median of 2,642 stars). Repositories owned by organizations have more stars than the ones owned by individuals.
\end{tcolorbox}\vspace{0.5em}

\vspace{0.5em}\noindent\emph{RQ \#2: Does stars correlate with repository's age, number of commits, number of contributors, and number of forks?}\vspace{0.5em}
\label{sub:results:rq3}

\begin{figure}[!ht]
\centering
\begin{subfigure}[b]{0.365\textwidth}
\includegraphics[width=\linewidth, trim={0 1em 0 4.5em}, clip]{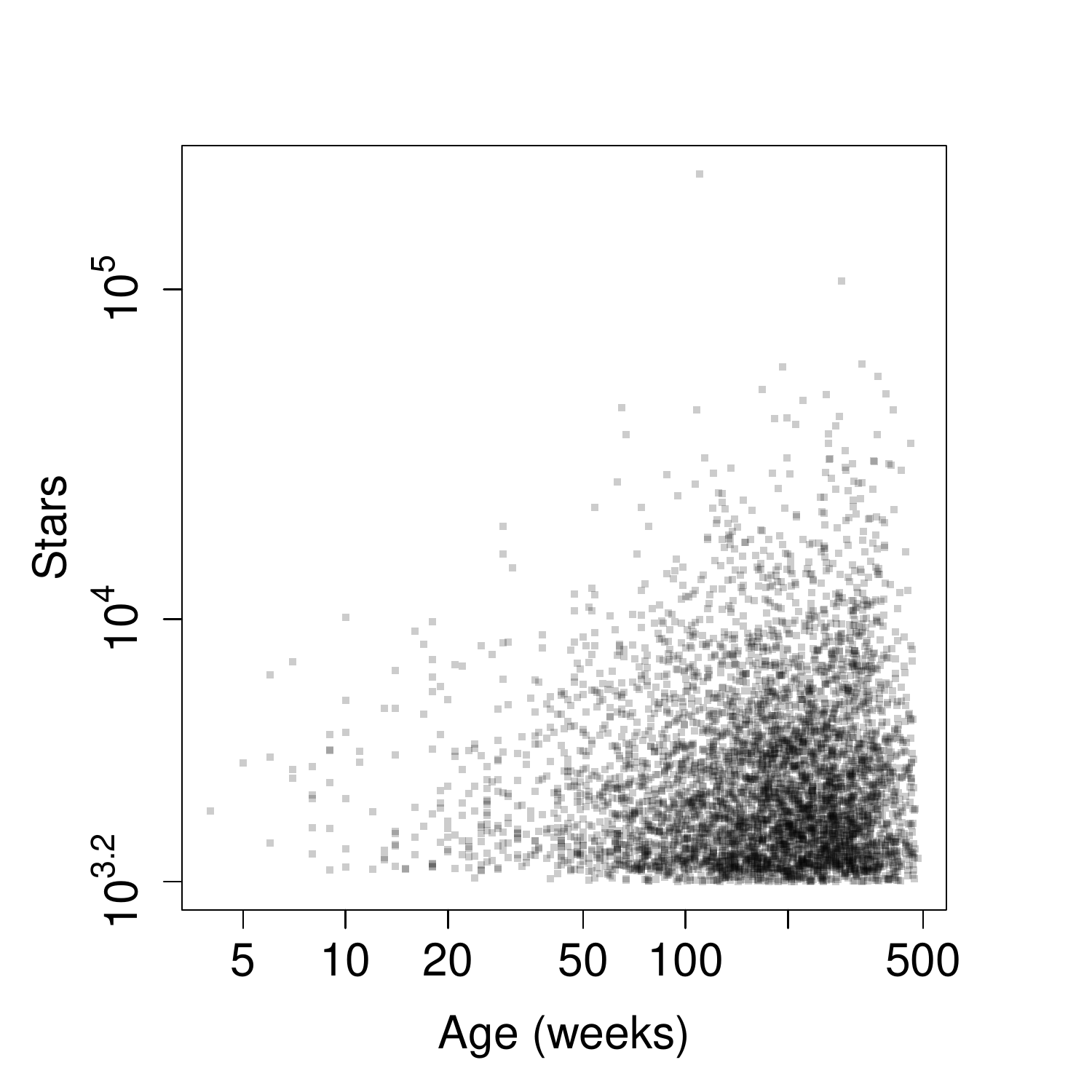}
\caption{Age vs Stars}
\label{fig:age-vs-stars}
\end{subfigure}%
\begin{subfigure}[b]{0.365\textwidth}
\includegraphics[width=\linewidth, trim={0 1em 0 4.5em}, clip]{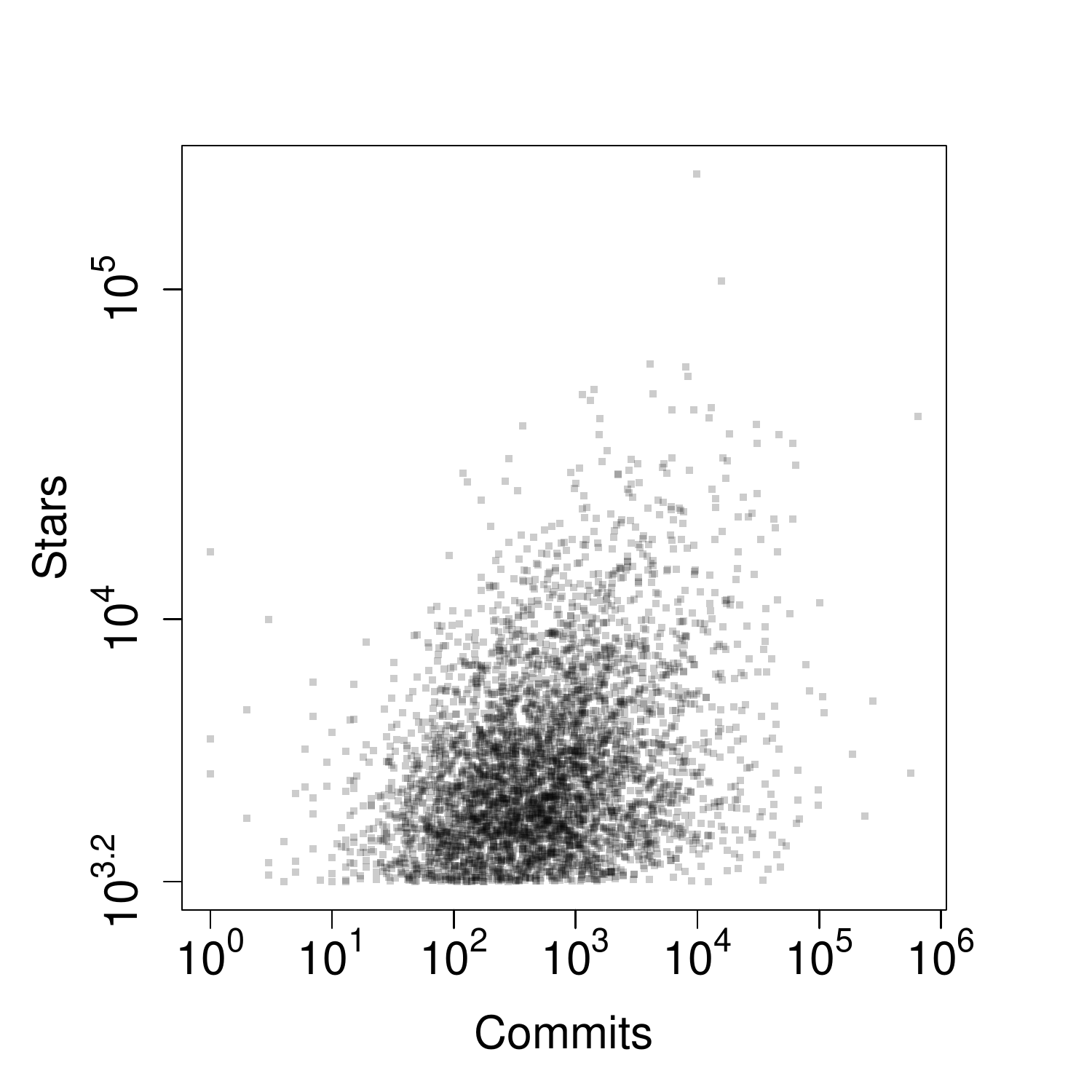}
\caption{Commits vs Stars}
\label{fig:commits-vs-stars}
\end{subfigure}%

\begin{subfigure}[b]{0.365\textwidth}
\includegraphics[width=\linewidth, trim={0 1em 0 4.5em}, clip]{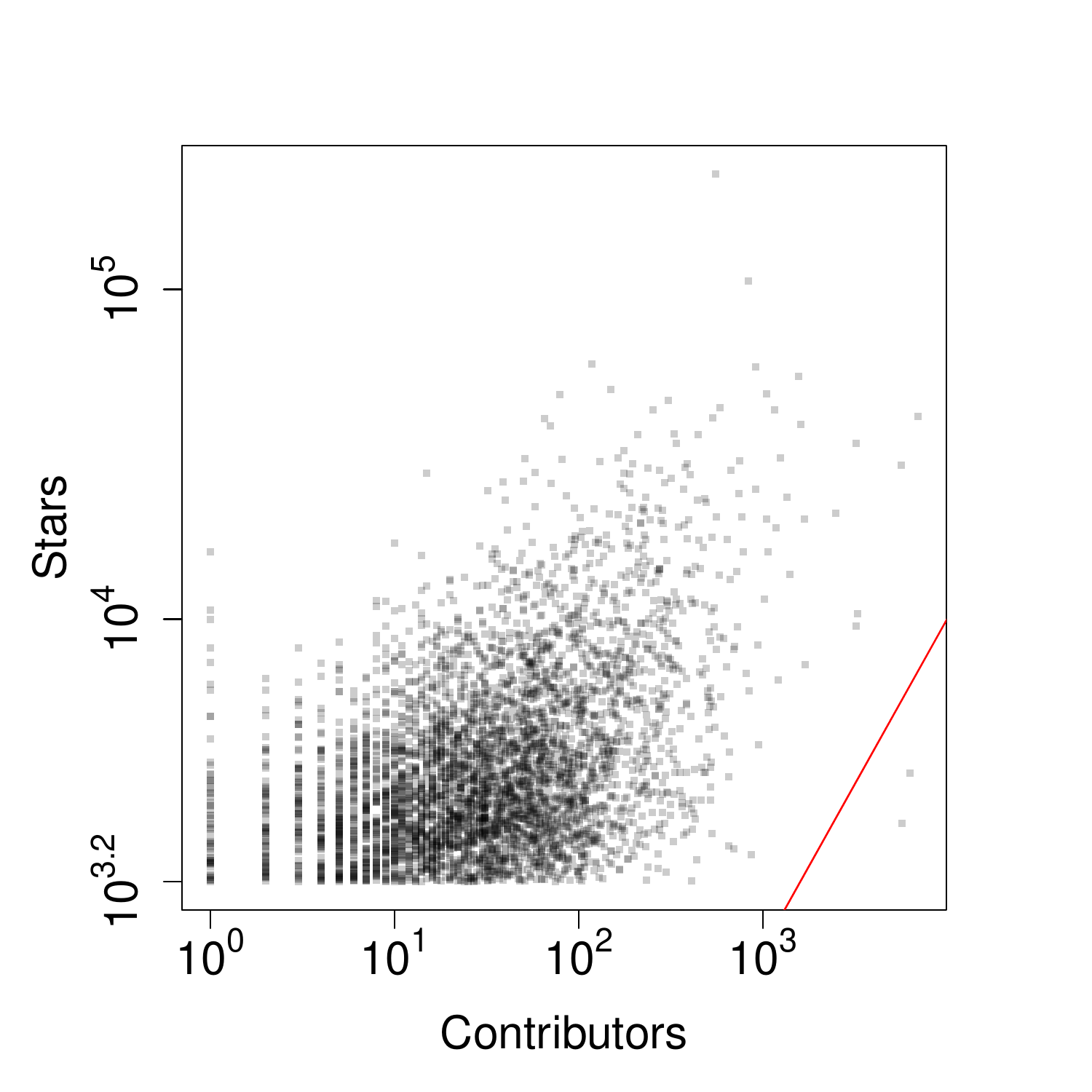}
\caption{Contributors vs Stars}
\label{fig:contributors-vs-stars}
\end{subfigure}%
\begin{subfigure}[b]{0.365\textwidth}
\includegraphics[width=\linewidth, trim={0 1em 0 4.5em}, clip]{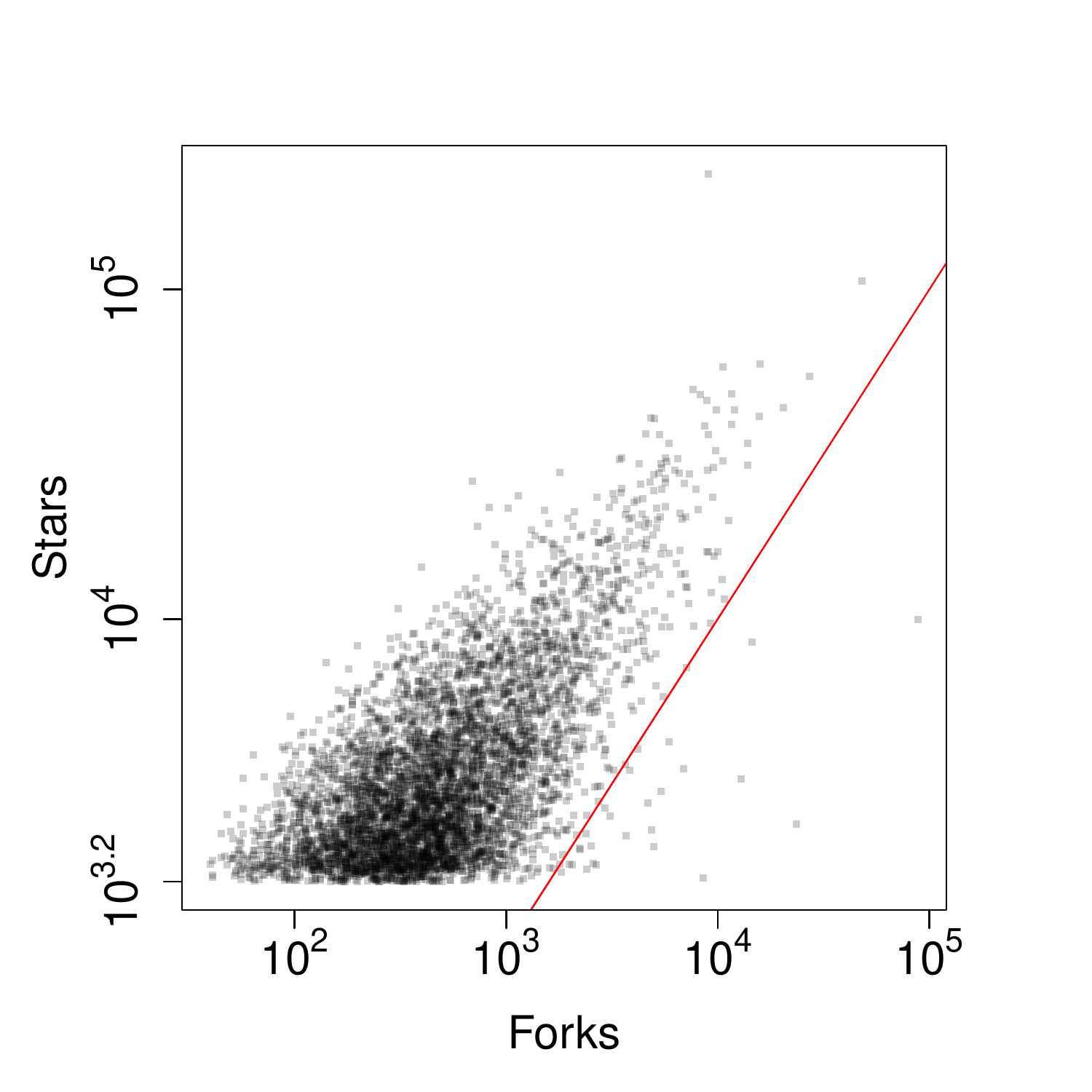}
\caption{Forks vs Stars}
\label{fig:forks-vs-stars}
\end{subfigure}
\caption{Correlation analysis. In subfigures (c) and (d), the line is the identity relation}
\label{fig:correlations}
\end{figure}

Figure~\ref{fig:correlations} shows scatterplots correlating the number of stars with the age (in number of weeks), number of commits, number of contributors, and number of forks of a repository.
Following the guidelines of Hinkle et al.~\cite{hinkle2003applied}, we interpret Spearman's rho as follows: 0.00 $\leqslant$ $rho$ $<$ 0.30 (negligible), 0.30 $\leqslant$ $rho$ $<$ 0.50 (low), 0.50 $\leqslant$ $rho$ $<$ 0.70 (moderate), 0.70 $\leqslant$ $rho$ $<$ 0.90 (high), and 0.90 $\leqslant$ $rho$ $<$ 1.00 (very high).
First, the plots suggest that stars are not correlated with the repository's age (Figure~\ref{fig:age-vs-stars}). We have old repositories with few stars and new repositories with many stars.
For example, \textsc{facebookincubator/create-react-app} has only five months and 19,083 stars, while \textsc{mojombo/grit} has more than 9 years and 1,883 stars.
Essentially, this result shows that repositories gain stars at different speeds.
We ran Spearman's rank correlation test and the resulting correlation coefficient is close to zero ($rho$ $=$ 0.050 and \emph{p-value} $<$ 0.001).

The scatterplot in Figure~\ref{fig:commits-vs-stars} shows that stars have a low correlation with number of commits ($rho$ $=$ 0.439 with \emph{p-value} $<$ 0.001).
However, as presented in Figure~\ref{fig:contributors-vs-stars}, stars have a moderate correlation with contributors ($rho$ $=$ 0.502 with \emph{p-value} $<$ 0.001).
In this figure, a logarithm scale is used in both axes; the line represents the identity relation: below the line are the systems with more contributors than stars.
Interestingly, two systems indeed have more contributors than stars: \textsc{raspberrypi/linux} (6,277 contributors and 3,414 stars) and \textsc{Linuxbrew/legacy-linuxbrew} (5,681 contributors and 2,397 stars).
This happens because they are forks of highly successful repositories (\textsc{torvalds/linux} and \textsc{Homebrew/brew}, respectively).
The top-3 systems with more stars per contributor are \textsc{shadowsocks/shadowsocks} (16,017 stars/ contributor), \textsc{wg/wrk} (10,658 stars/contributor), and \textsc{octocat/Spoon-Knife} (9,961 stars/contributor). However, these systems have just one contributor.
The three systems with less stars per contributor are \textsc{DefinitelyTyped/DefinitelyTyped} (2.97 stars/contributor), \textsc{nodejs/node-convergence-archive} (2.88 stars/contributor), and \textsc{openstack/nova} (2.23 stars/contributor).

Finally, Figure~\ref{fig:forks-vs-stars} shows plots correlating stars and forks.
As suggested by the followed guidelines, there is a moderate positive correlation between the two measures ($rho$ $=$ 0.558 and \emph{p-value} $<$ 0.001).
For example, \textsc{twbs/bootstrap} is the second repository with the highest number of stars and also the second one with more forks. \textsc{angular/angular.js} is the fifth repository in number of stars and the third one with more forks.
In Figure~\ref{fig:forks-vs-stars}, we can also observe that only 28 systems (0.56\%) have more forks than stars.
As examples, we have a repository that just provides a tutorial for forking a repository (\textsc{octocat/SpoonKnife}) and a popular puzzle game (\textsc{gabrielecirulli/2048}), whose success motivated many forks with variations of the original implementation.

\vspace{2em}\begin{tcolorbox}[left=0.75em,right=0.75em,top=0.75em,bottom=0.75em,boxrule=0.25mm,colback=gray!5!white]
\noindent {\em Summary:} There is no correlation between stars and repository's age; however, there is a low correlation with commits, and a moderate correlation with contributors and forks. 
\end{tcolorbox}\vspace{2em}

\vspace{0.5em}\noindent\emph{RQ \#3: How early do repositories get their stars?}\vspace{0.5em}
\label{sub:results:rq4}

\begin{figure}[!ht]
\centering
\includegraphics[width=0.45\linewidth, page=1, trim={0 0 0 0}, clip]{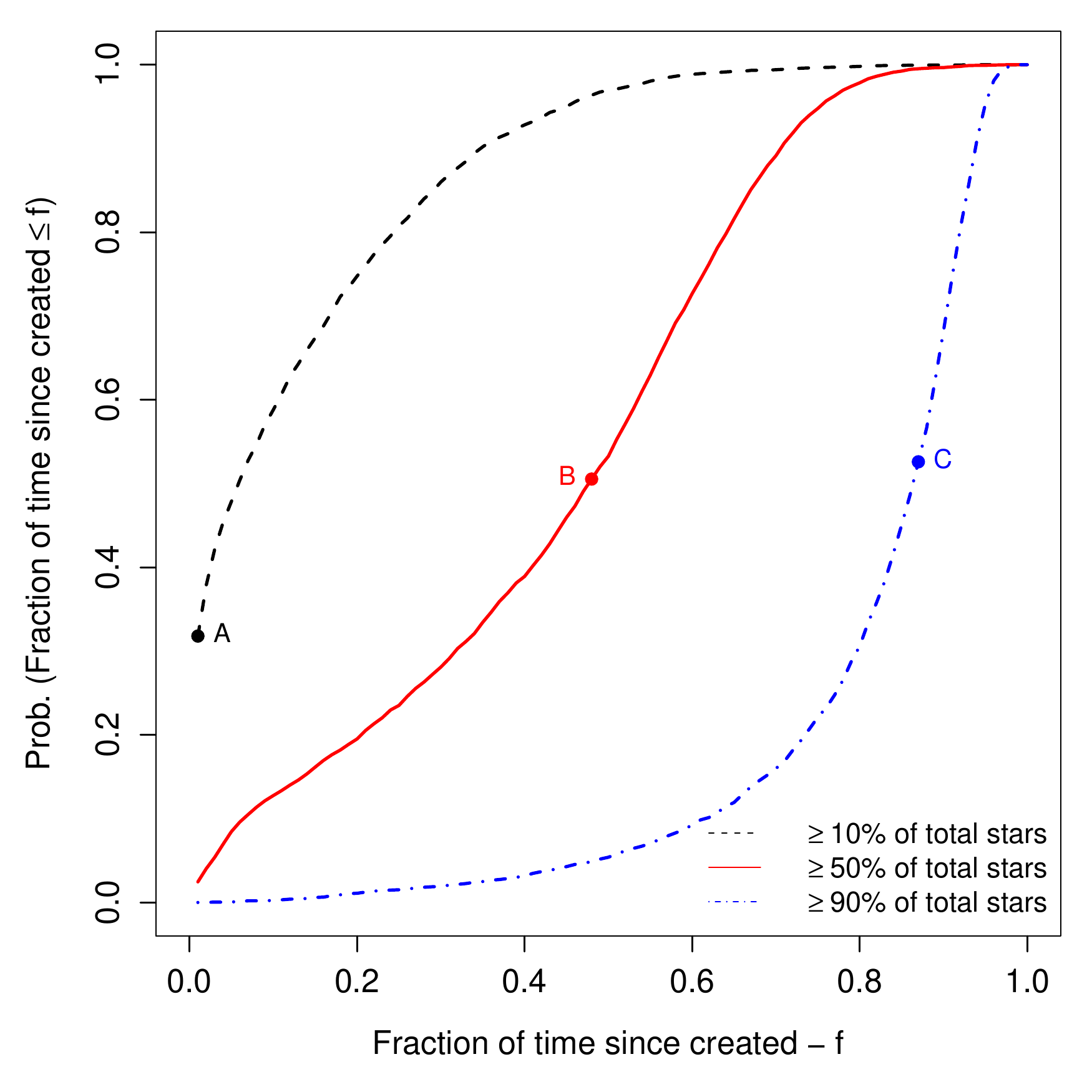}
\caption{Cumulative distribution of the time fraction a repository takes to receive 10\%, 50\%, and 90\% of its stars}
\label{fig:cdf}
\end{figure}

Figure~\ref{fig:cdf} shows the cumulative distribution of the time fraction a repository takes to receive at least 10\%, at least 50\%, and at least 90\% of its stars.
Around 32\% of the repositories receive 10\% of their stars very early, in the first days after the initial release (label A, in Figure~\ref{fig:cdf}).
We hypothesize that many of these initial stars come from early adopters, who start commenting and using novel open source software immediately after they are public released.\footnote{\label{footnote:release}It is worth mentioning that GitHub repositories can be created private and turned public later. In this RQ, we consider the latter event, which we referred as public release.}
After this initial burst, the growth of the number of stars tend to stabilize.
For example, half of the repositories take 48\% of their age to receive 50\% of their stars (label B); and around half of the repositories take 87\% of their age to receive 90\% of their number of stars (label C).

\begin{figure}[!ht]
\centering
\includegraphics[width=0.4\linewidth, page=1, trim={0 3em 0 0}, clip]{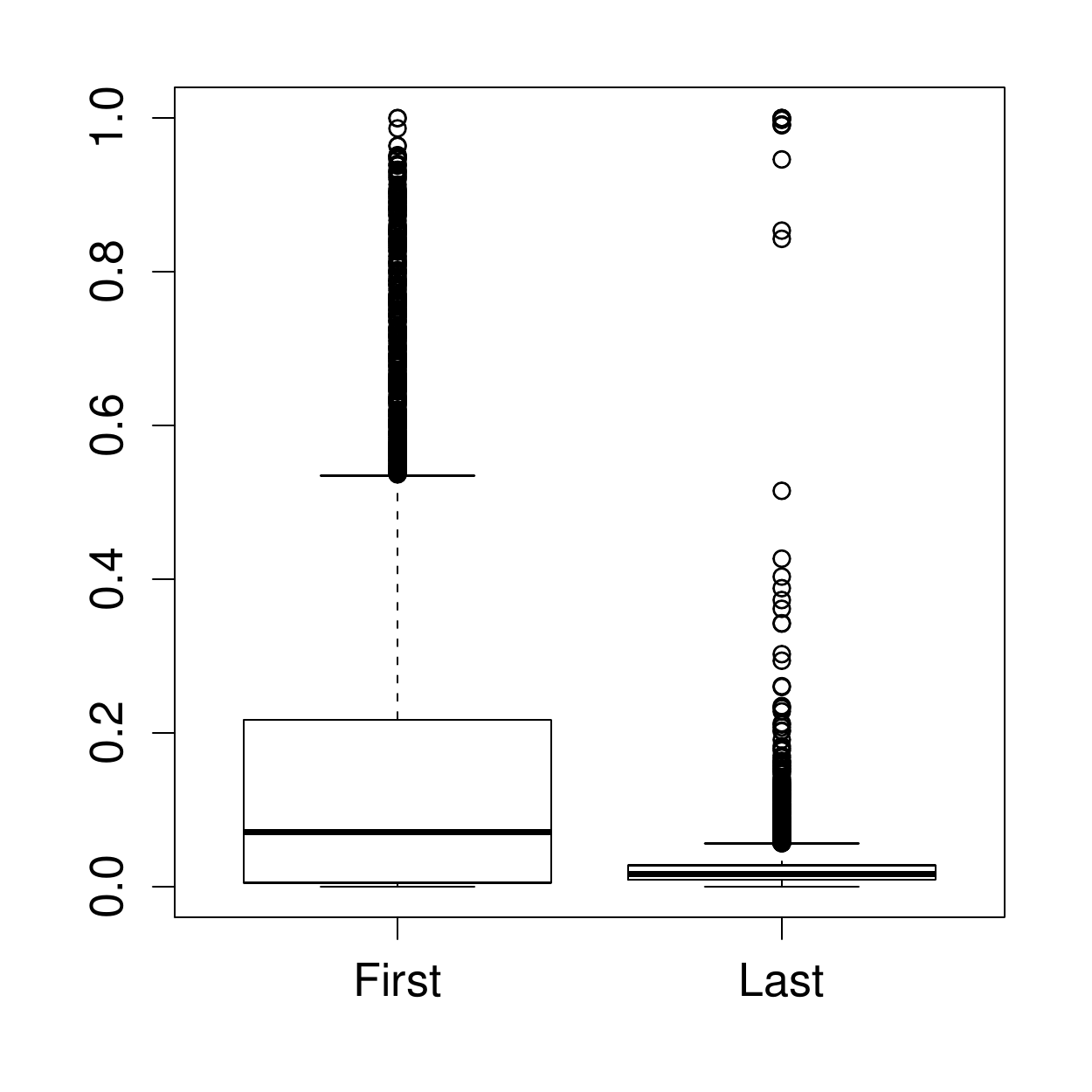}
\caption{Fraction of stars gained in the first four weeks and in the last four weeks}
\label{fig:cdf_stages}
\end{figure}

Figure~\ref{fig:cdf_stages} shows the distribution of the fraction of stars gained in the first and last four weeks of the repositories.
For the first four weeks, the fraction of stars gained is 0.4\% (first quartile), 7.0\% (second quartile), and 21.6\% (third quartile).
For the last four weeks, it is 0.8\% (first quartile), 1.6\% (second quartile), and 2.7\% (third quartile). By applying the Mann-Whitney test, we found that these distributions are different (\emph{p-value} $<$ 0.001) with a {\em large} effect size (Cohen's $d = 0.856$).

\vspace{1em}\begin{tcolorbox}[left=0.75em,right=0.75em,top=0.75em,bottom=0.75em,boxrule=0.25mm,colback=gray!5!white]
\noindent{\em Summary:} Repositories have a tendency to receive more stars right after their public release. After that, the growth rate tends to stabilize. 
\end{tcolorbox}\vspace{0.5em}

\vspace{0.5em}\noindent\emph{RQ \#4: What is the impact of new features on stars?}\vspace{0.5em}
\label{sub:results:rq5}

In this research question, we investigate the impact of new features on the number of stars of GitHub repositories.
The goal is to check whether the implementation of new features (resulting in new releases of the projects) contribute to a boost in the number of stars.
Specifically, we selected 1,539 repositories from our dataset (30.7\%) that follow a semantic versioning convention to number releases.
In such systems, versions are identified by three integers, in the format $x.y.z$, with the following semantics: increments in ${x}$ denote major releases, which can be incompatible with older versions; increments in $y$ denote minor releases, which add functionality in a backward-compatible manner; and increments in $z$ denote bug fixes. In our sample, we identified 1,304 major releases and 8,570 minor releases.

\begin{figure}[!ht]
\centering
\includegraphics[width=0.55\linewidth, page=1, trim={0 1.5em 0 2.5em}, clip]{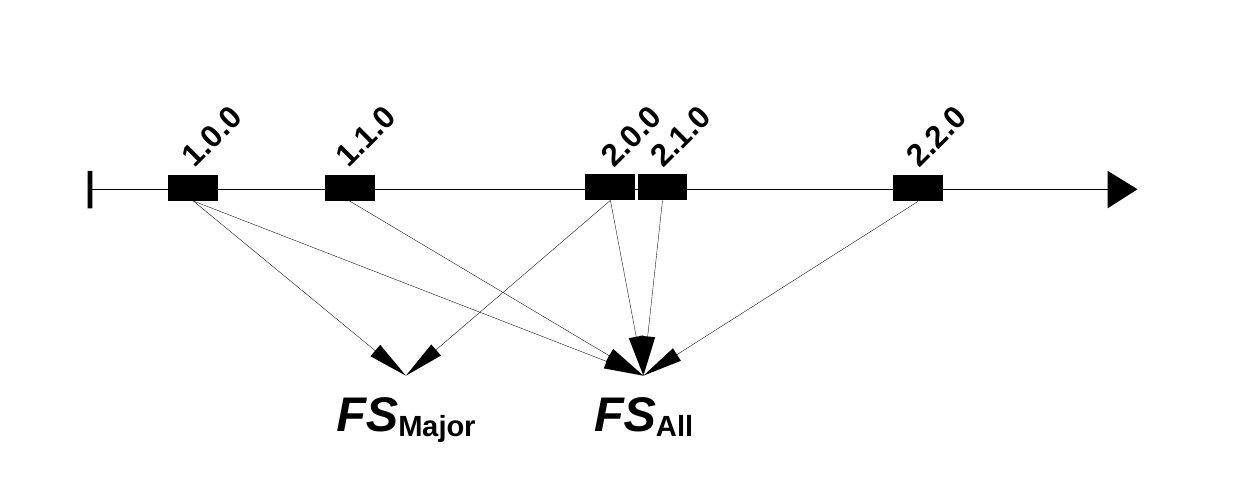}
\caption{Fraction of stars for all releases ($\mathit{FS}\textsubscript{All}$) and just after major releases ($\mathit{FS}\textsubscript{Major}$)}
\label{fig:releases_schema}
\end{figure}

\begin{figure}[!ht]
\centering
\includegraphics[width=0.4\linewidth, page=1, trim={0 1.25em 0 1.5em}, clip]{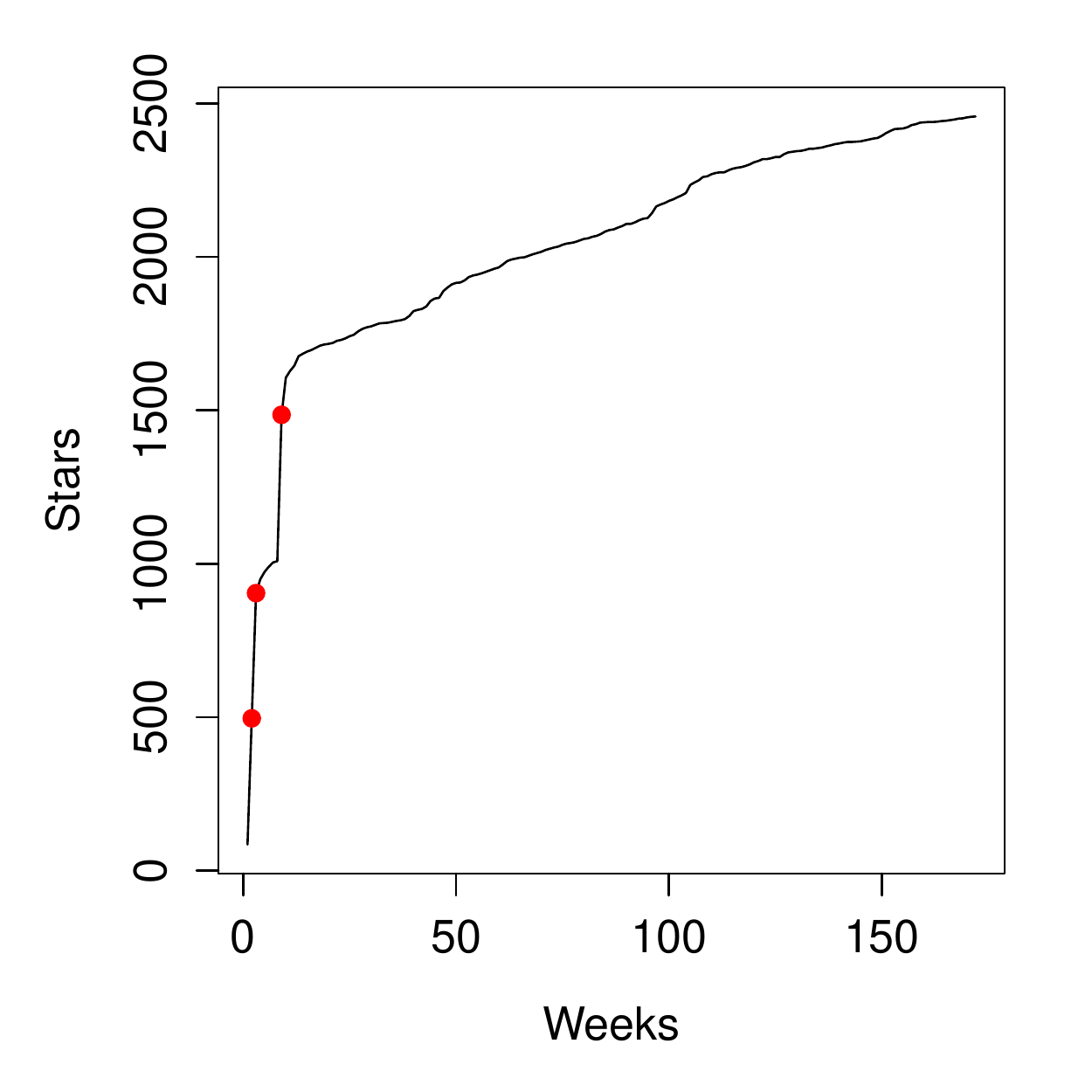}
\caption{\textsc{Reportr/dashboard} (the dots indicate weeks with releases)}
\label{fig:releases2example}
\end{figure}

First, as illustrated in Figure~\ref{fig:releases_schema}, we counted the fraction of stars received by each repository in the week following all releases ($\mathit{FS}\textsubscript{All}$) and just after major releases ($\mathit{FS}\textsubscript{Major}$).
As mentioned, the goal is to check the impact of new features in the number of stars right after new releases
(however, in the end of the RQ, we also consider the impact of different week intervals).
As an example, Figure~\ref{fig:releases2example} shows the time series for \textsc{Reportr/dashboard}, using dots to indicate the project's releases (v1.0.0/v.1.1.0, v2.0.0, and v2.1.0, respectively).
This project has $FS\textsubscript{All} = 0.525$ (i.e., 52.5\% of its stars were gained in the weeks following the four releases) and $FS\textsubscript{Major} = 0.248$  (i.e., 24.8\% of its stars were gained in the weeks following the releases v1.0.0 and v2.0.0).

\begin{figure}[!ht]
\centering
\includegraphics[width=0.4\linewidth, page=1, trim={0 1.25em 0 1.5em}, clip]{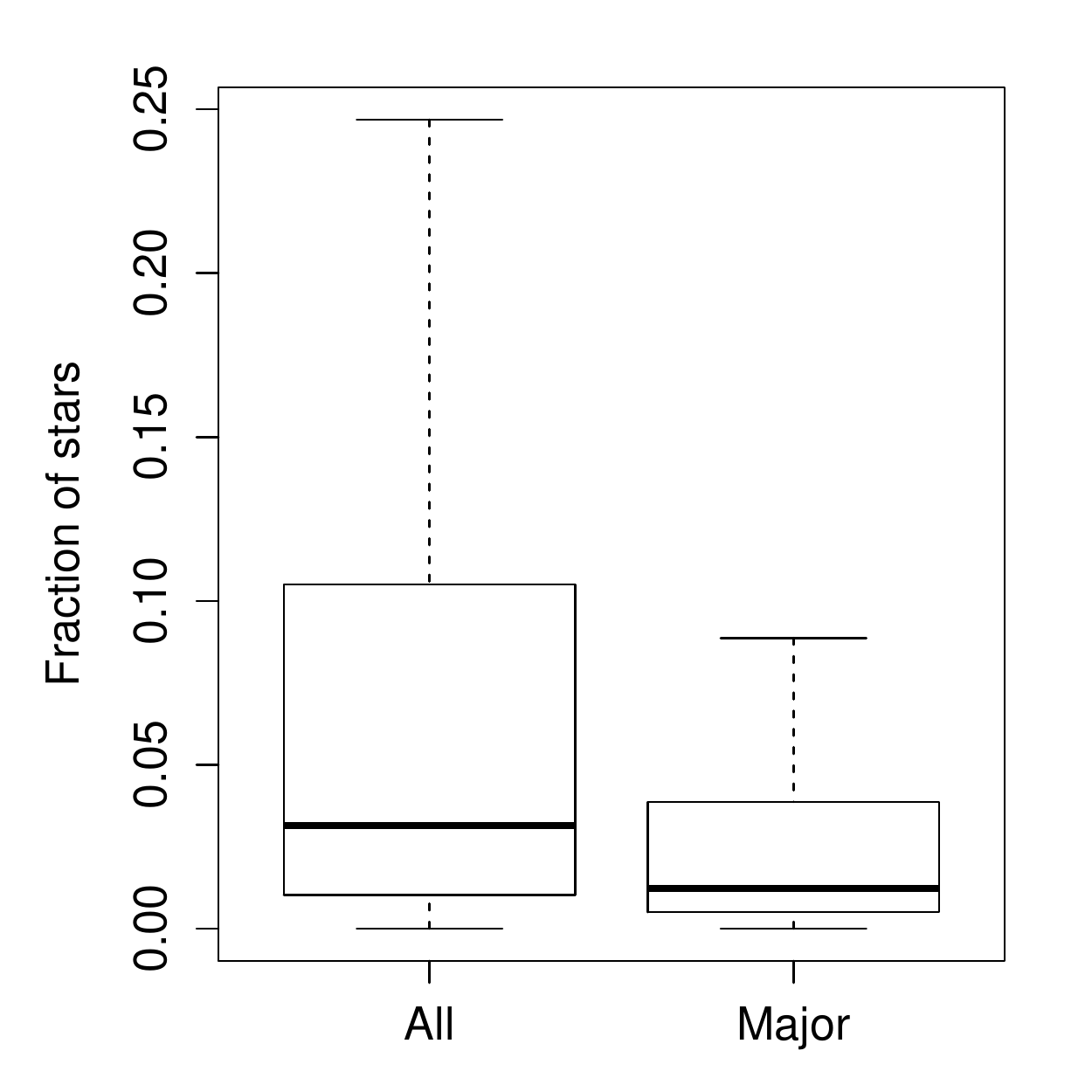}
\caption{Fraction of stars gained in the first week after all releases and just after the major releases}
\label{fig:releases1}
\end{figure}

Figure~\ref{fig:releases1} shows the distribution of $\mathit{FS}\textsubscript{All}$ and $\mathit{FS}\textsubscript{Major}$ for all selected repositories.
When considering all releases, the fraction of stars gained in the first week after the releases is 1.0\% (first quartile), 3.1\% (second quartile), and 10.5\% (third quartile). For the major releases, it is 0.5\% (first quartile), 1.2\% (second quartile), and 3.8\% (third quartile). By applying the Mann-Whitney test, we found that these distributions are different (\emph{p-value} $<$ 0.001), but with a {\em small} effect size (Cohen's $d = 0.316$).
\textsc{yarnpkg/yarn} (a package manager for JavaScript) is the repository with the highest fraction of stars received after releases. The repository has one year, 21,809 stars, and gained most of its stars (83.0\%) in the weeks after its releases.

\begin{figure}[!ht]
\centering
\includegraphics[width=0.4\linewidth, page=3, trim={0 1.25em 0 2.5em}, clip]{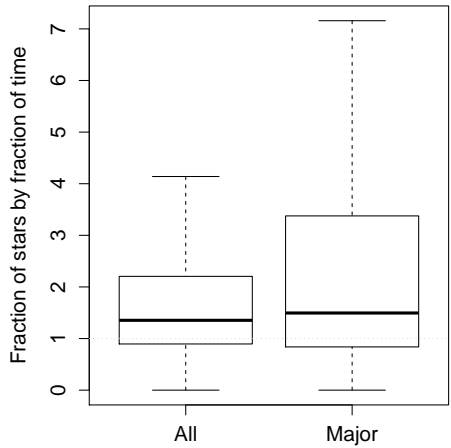}
\caption{Fraction of stars gained in the week following all releases (or just the major releases) $/$ fraction of time represented by these weeks}
\label{fig:releases2}
\end{figure}

We also computed two ratios: $\mathit{R}\textsubscript{All} = \mathit{FS}\textsubscript{All}/\mathit{FT}\textsubscript{All}$ and $\mathit{R}\textsubscript{Major} = \mathit{FS}\textsubscript{Major}/\mathit{FT}\textsubscript{Major}$, where $\mathit{FT}$ is the fraction of time represented by the weeks following the releases per the repository's age.
When $\mathit{R}\textsubscript{All} > 1$ or $\mathit{R}\textsubscript{Major} > 1$, the repository gains proportionally more stars after releases.
For example, \textsc{Reportr/dashboard} (Figure~\ref{fig:releases2example}) has $\mathit{FT}\textsubscript{All} = 0.019$ (i.e., the weeks following all releases represent only 1.9\% of its total age) resulting in $\mathit{R}\textsubscript{All} = 0.525 / 0.019 = 27.047$.
Therefore, releases have a major impact on its number of stars.
Figure~\ref{fig:releases2} shows boxplots with the results of $\mathit{R}\textsubscript{All}$ and $\mathit{R}\textsubscript{Major}$ for all repositories.
Considering all releases, we have that $\mathit{R}\textsubscript{All}$ is 0.89 (first quartile), 1.35 (second quartile), and 2.20 (third quartile). For major releases only, we have that $\mathit{R}\textsubscript{Major}$ is 0.83 (first quartile), 1.49 (second quartile), and 3.37 (third quartile). By applying the Mann-Whitney test, we found that these distributions are different (\emph{p-value} $<$ 0.05); but after computing Cohen's $d$, we found a {\em very small} effect size ($d = -0.188$).

\begin{figure}[!ht]
\centering
\includegraphics[width=0.4\linewidth, trim={0 1.5em 0 2.5em}, clip]{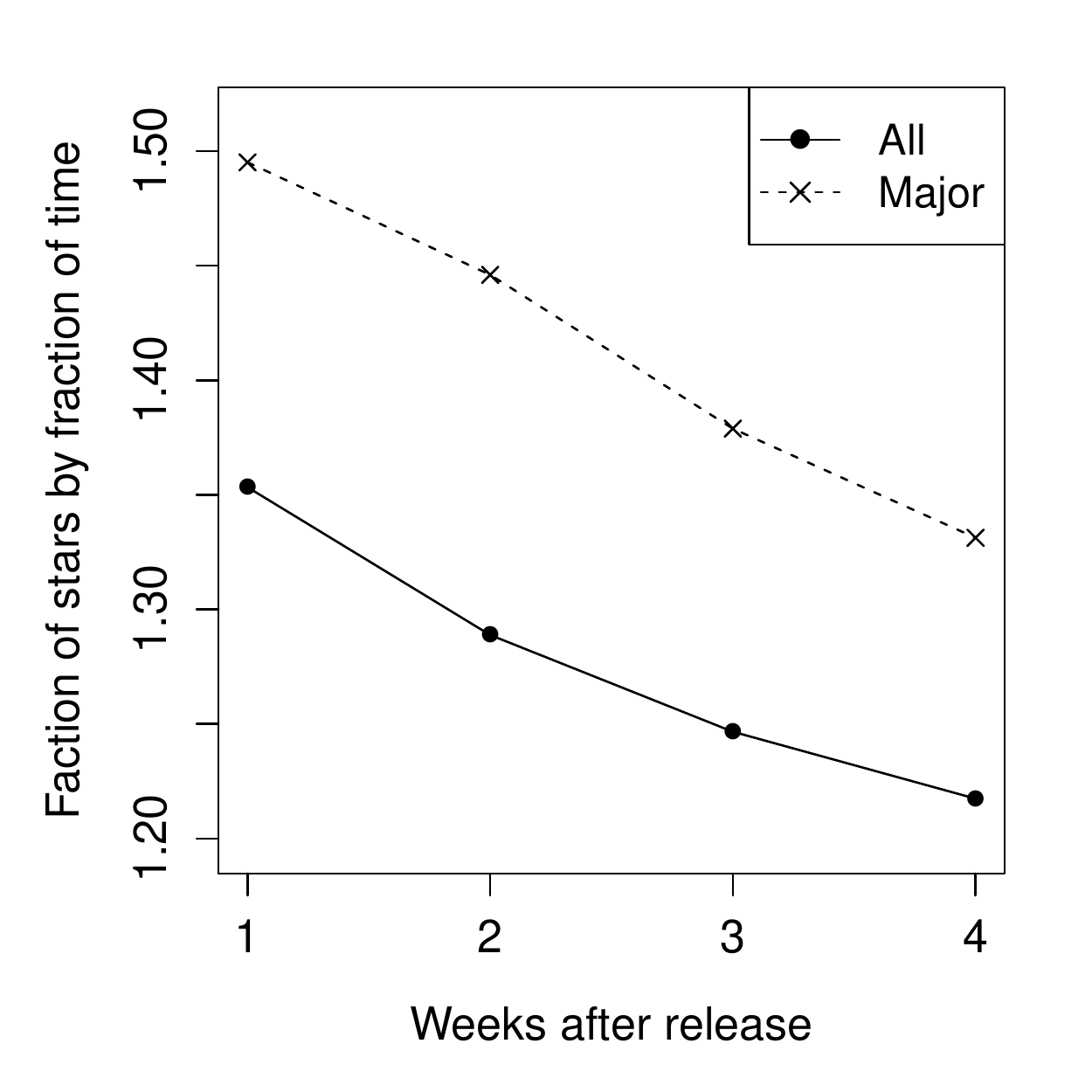}
\caption{Fraction of stars by fraction of time (median values), computed using different time intervals}
\label{fig:releases-values}
\end{figure}

Finally, Figure~\ref{fig:releases-values} shows the median values of $\mathit{R}\textsubscript{All}$ and $\mathit{R}\textsubscript{Major}$ computed using stars gained after $n$ weeks ($1 \leq n \leq 4$).
Both ratios decrease (for major and all releases).
Therefore, although there is some gains of stars after releases, they tend to decrease after few weeks.

\vspace{1em}\begin{tcolorbox}[left=0.75em,right=0.75em,top=0.75em,bottom=0.75em,boxrule=0.25mm,colback=gray!5!white]
\noindent{\em Summary:} There is an acceleration in the number of stars gained after releases. For example, half of the repositories gain at least 49\% more stars in the week following major releases than in other weeks. However, because repositories usually have more weeks without releases, this phenomenon is not sufficient to result in a major concentration of stars after releases. For example, 75\% of the systems gain at most 3.8\% of their stars in the week following major releases.
\end{tcolorbox}\vspace{0.5em}

\noindent {\em Implications for Empirical Software Engineering Researchers:}
Regarding the selection of GitHub projects for empirical studies based on number of stars, the following observations are derived from our findings: (1)
this selection favors JavaScript systems  (31.1\% of the systems in our dataset) and also web libraries and frameworks (30.7\% of the dataset systems); (2) this selection might result in a relevant number of projects that are not software systems (8.6\% of the projects in our dataset are tutorials, books, {\em awesome-lists}, etc);
(3) this selection favors large projects (in terms of number of contributors) with many forks, as we concluded when investigating RQ \#2 (correlation analysis); (4) additionally, after examining RQ \#3, we recommend researchers (and also practitioners) to check whether the stars are not gained in a short time interval, for example, after the  project public release.

% !TEX root = jss.tex
\section{Stars Growth Patterns}
\label{sec:patterns}

In this section, we investigate common growth patterns concerning the number of stars of the GitHub repositories in our dataset. To this purpose, we use the KSC algorithm~\cite{Yang2011}. 
This algorithm uses an iterative approach, similar to the classical K-means clustering algorithm, to assign the time series in clusters and then refine the clusters centroids by optimizing a specific time series distance metric that is invariant to scaling and shifting. As result, the clusters produced by the KSC algorithm are less influenced by outliers.
KSC is used in other studies to cluster time series representing the popularity of YouTube videos~\cite{Figueiredo2013} and Twitter posts~\cite{Lehmann2012}. Like K-means~\cite{hartigan1975}, KSC requires as input the number of clusters $k$.

Because the time series provided as input to KSC must have the same length, we only consider data regarding the last 52 weeks (one year). We acknowledge that this decision implies a comparison of projects in different stages of their evolution (e.g.,~a very young project, which just completed one year, and mature projects, with several years). However, it guarantees the derivation of growth patterns explaining the dynamics of the most recent stars received by a project and in this way it also increases the chances of receiving valuable feedback of the projects contributors, in the survey described in Section~\ref{sec:patterns-survey}.
Due to this decision, we had to exclude from our analysis 333 repositories (6.6\%) that have less than 52 weeks.

\begin{figure}[!ht]
\centering
\includegraphics[width=0.4\linewidth, trim={0 1.5em 0 2.5em}, clip]{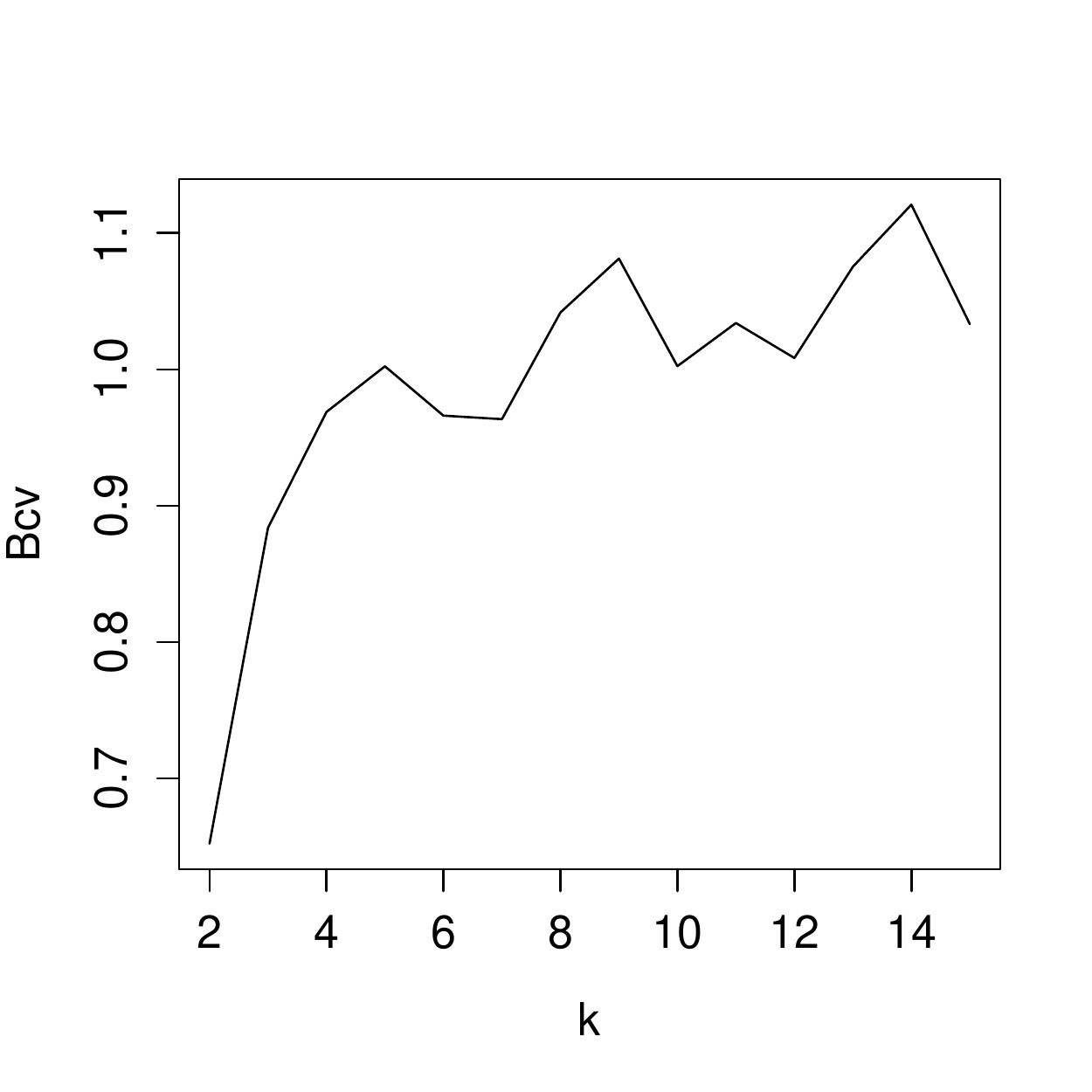}
\caption{$\beta_{CV}$ for $2 \leq k \leq 15$}
\label{fig:Bcv}
\end{figure}

We use the $\beta_{CV}$ heuristic~\cite{Menasce2001} to define the best number $k$ of clusters. $\beta_{CV}$ is defined as the ratio of two coefficients: variation of the intracluster distances and variation of the intercluster distances.
The smallest value of $k$ after which the $\beta_{CV}$ ratio remains roughly stable should be selected. This means that new added clusters affect only marginally the intra and intercluster variations~\cite{Figueiredo2014}.
In our dataset, the values of $\beta_{CV}$ start to stabilize for $k=4$ (see Figure~\ref{fig:Bcv}).
Note that although the value of $\beta_{CV}$ increases for $k=5$ (from 0.968 to 1.002, respectively), the $\beta_{CV}$ for  $k=4$ remains almost the same for $k=6$ and $k=7$ (0.966 and 0.963, respectively). For this reason, we use four clusters in this study.

\subsection{Proposed Growth Patterns}
\label{sec:patterns:trends}

\begin{figure}[!ht]
\centering
\begin{subfigure}[b]{0.325\textwidth}
\includegraphics[width=\linewidth, trim={0 1.5em 2em 3.5em}, clip]{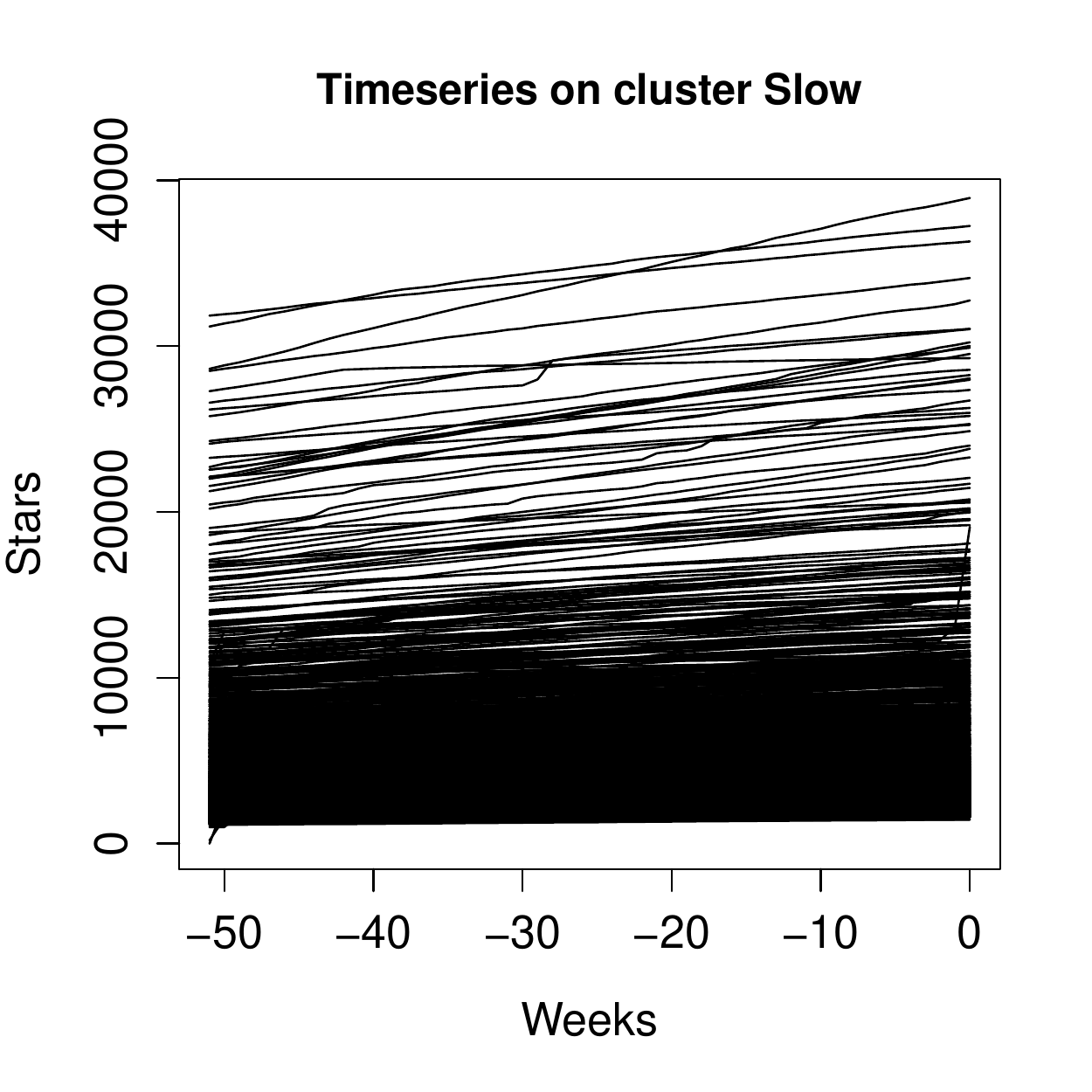}
\caption{Cluster C1}
\label{fig:timeseries:c1}
\end{subfigure}%
\begin{subfigure}[b]{0.325\textwidth}
\includegraphics[width=\linewidth, trim={0 1.5em 2em 3.5em}, clip]{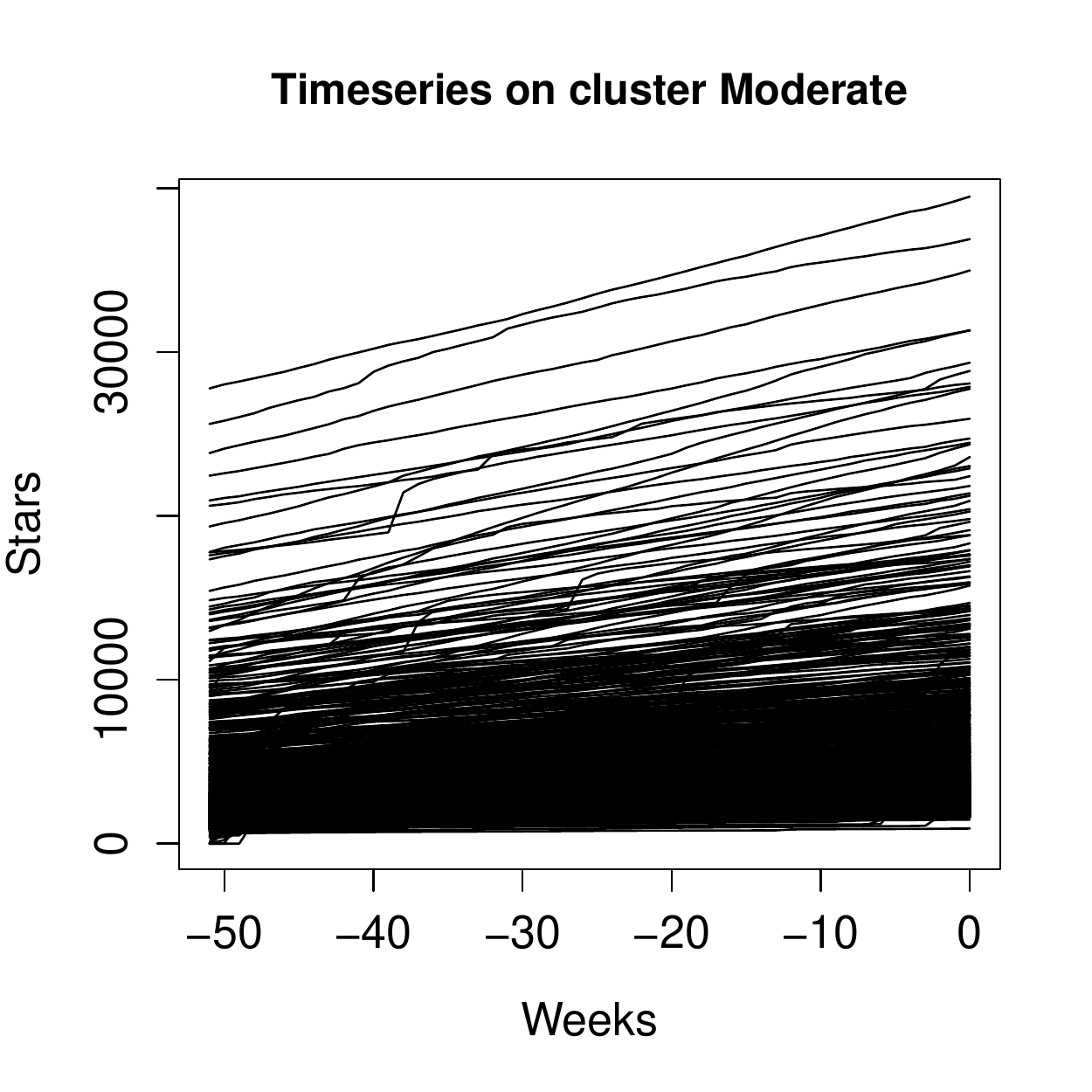}
\caption{Cluster C2}
\label{fig:timeseries:c2}
\end{subfigure}%

\begin{subfigure}[b]{0.325\textwidth}
\includegraphics[width=\linewidth, trim={0 1.5em 2em 3.5em}, clip]{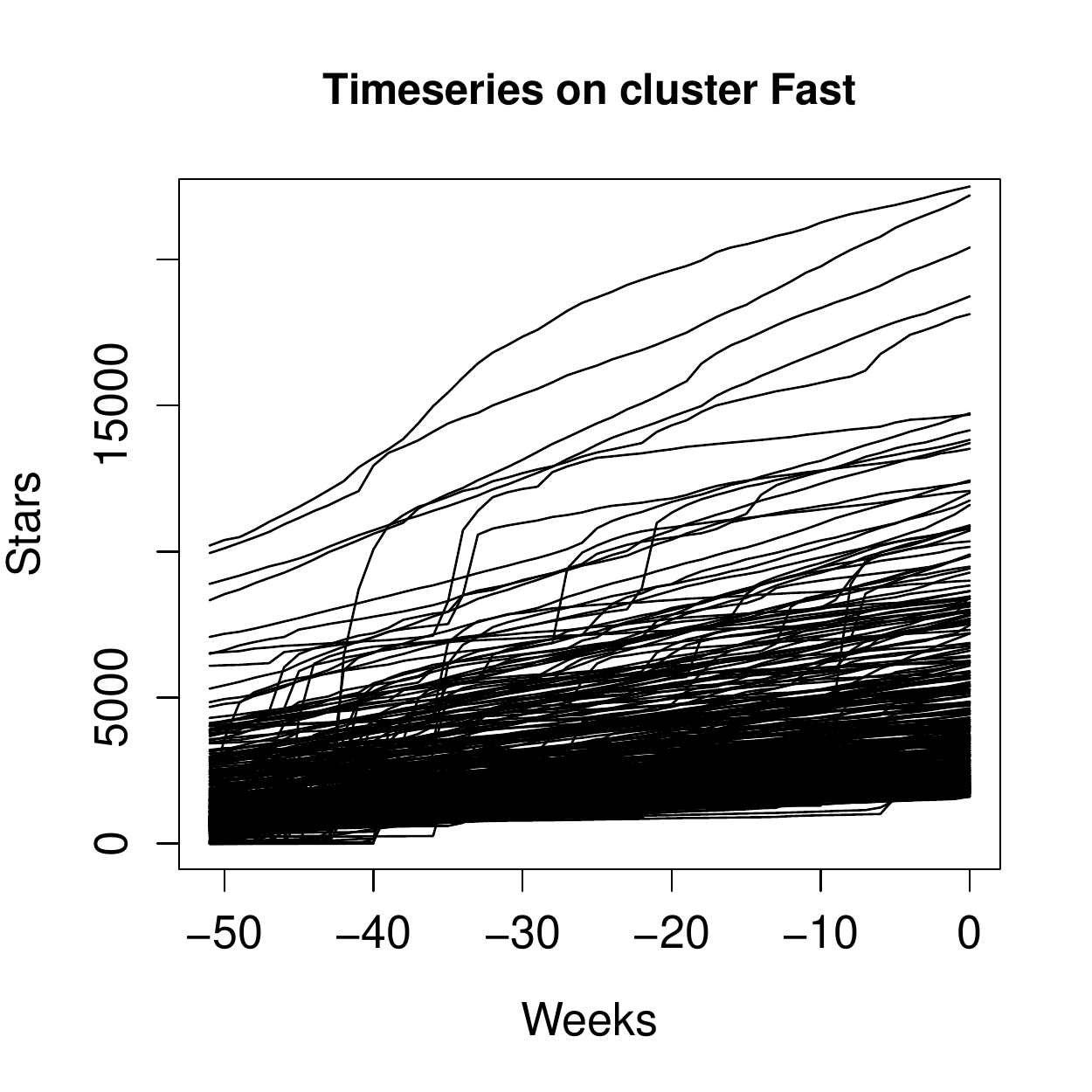}
\caption{Cluster C3}
\label{fig:timeseries:c3}
\end{subfigure}%
\begin{subfigure}[b]{0.325\textwidth}
\includegraphics[width=\linewidth, trim={0 1.5em 2em 3.5em}, clip]{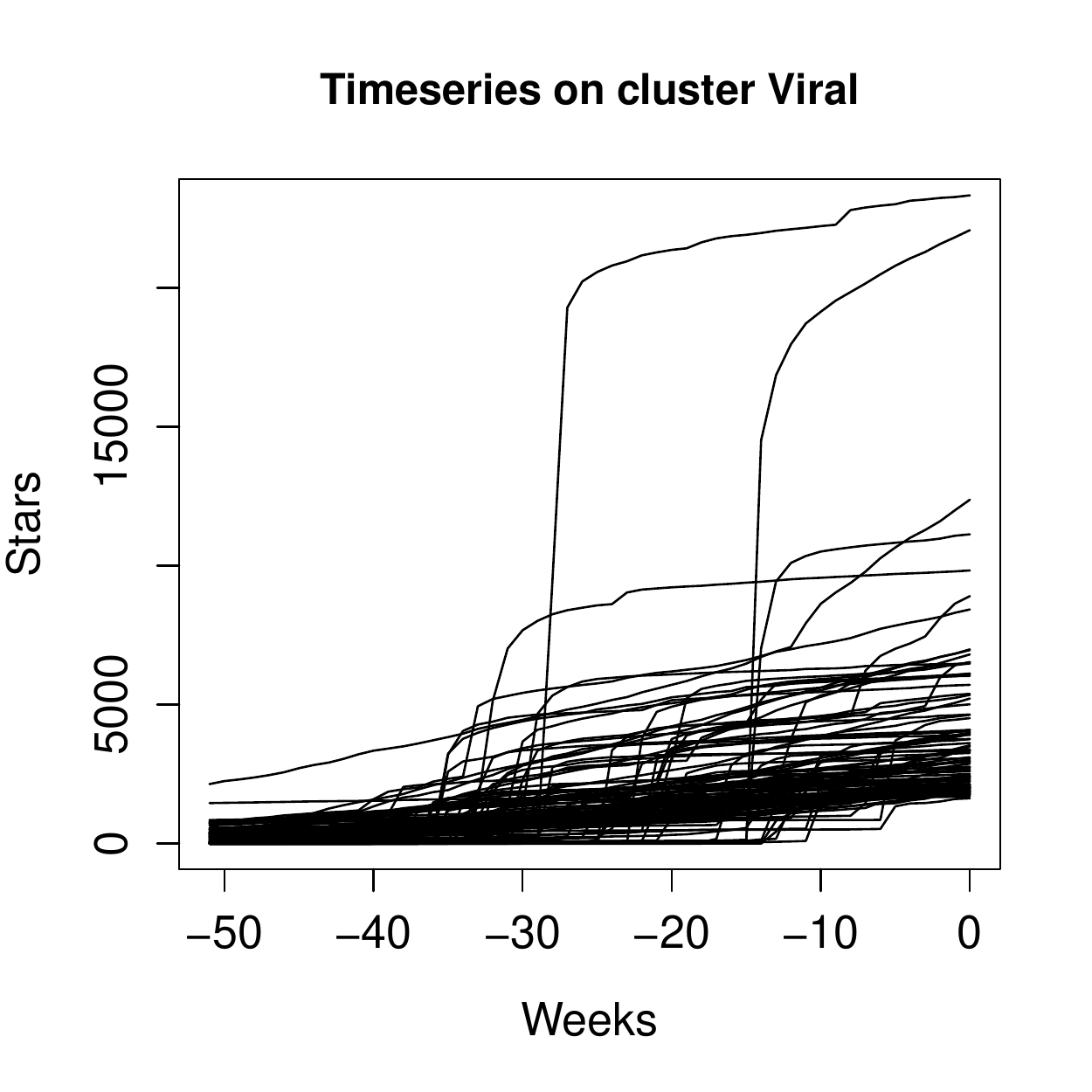}
\caption{Cluster C4}
\label{fig:timeseries:c4}
\end{subfigure}%
\caption{Clusters of time series produced by the KSC algorithm}
\label{fig:timeseries}
\end{figure}

\begin{figure}[!ht]
\centering
\includegraphics[width=0.55\columnwidth, trim={0 0.5em 0 0}, clip]{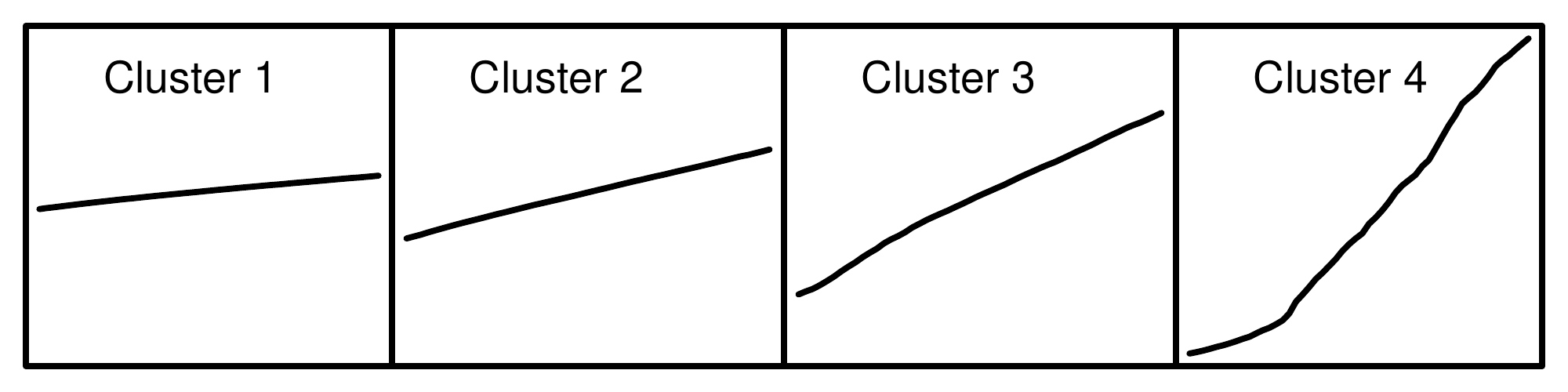}
\caption{Time series representing the centroids of each cluster}
\label{fig:popularity-trends}
\end{figure}

Figure~\ref{fig:timeseries} shows plots with the time series in each cluster. The time series representing the clusters' centroids are presented in Figure~\ref{fig:popularity-trends}.
The time series in clusters C1, C2, and C3 suggest a linear growth, but at different speeds. On the other hand, the series in cluster C4 suggest repositories with a sudden growth in the number of stars. We refer to these clusters as including systems with \emph{Slow}, \emph{Moderate}, \emph{Fast}, and \emph{Viral} Growth, respectively.

\begin{figure}[!ht]
\centering
\begin{subfigure}[t]{0.4\textwidth}
\centering
\includegraphics[width=1\linewidth, page=1, trim={0 1em 0 4em}, clip]{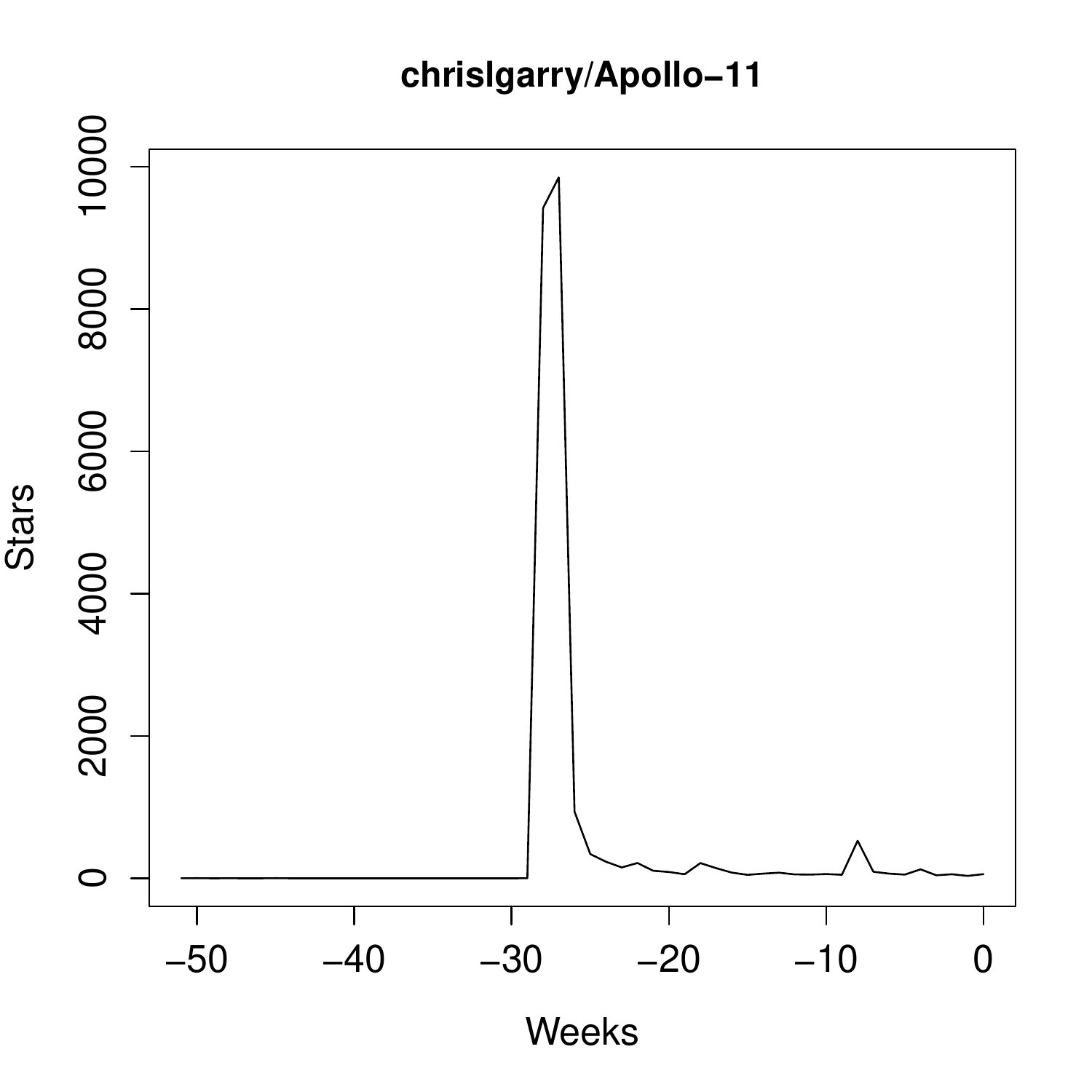}
\caption{\textsc{Apollo-11}}
\label{fig:viral-examples:a}
\end{subfigure}%
\begin{subfigure}[t]{0.4\textwidth}
\centering
\includegraphics[width=1\linewidth, page=4, trim={0 1em 0 4em}, clip]{img/timeseries_viral_single.pdf}
\caption{\textsc{Tesseract.js}}
\label{fig:viral-examples:b}
\end{subfigure}%
\caption{Examples of systems with viral growth}
\label{fig:viral-examples}
\end{figure}

Slow growth is the dominant pattern, including 58.2\% of the repositories in our sample, as presented in Table~\ref{tab:linear}. The table also shows the percentage of stars gained by the cluster's centroids in the period under analysis (52 weeks). The speed in which the repositories gain stars in cluster C1 is the lowest one (19.8\% of new stars in one year). Moderate growth is the second pattern with more repositories (30.0\% of the repositories and 63.9\% of new stars in one year). 9.3\% of the repositories have a fast growth (218.6\% of new stars in the analyzed year).
Cluster C4 (Viral Growth) includes repositories with a massive growth in their number of stars (1,317\%). However, it is a less common pattern, including 2.3\% of the repositories. Figure~\ref{fig:viral-examples} shows two examples of systems with a viral growth: \textsc{chrislgarry/Apollo--11} (Apollo 11 guidance computer source code, with a peak of 19,270 stars in two weeks) and \textsc{naptha/tesseract.js} (a JavaScript library to recognize words in images, which received 6,888 stars in a single week). 

\begin{table}[!ht]
\centering
\caption{Stars Growth Patterns}
\label{tab:linear}
\begin{tabular}{@{}ccrr@{}}
  \toprule
  \multicolumn{1}{c}{Cluster} & \multicolumn{1}{c}{Pattern} & \multicolumn{1}{c}{\# Repositories} & \multicolumn{1}{c}{Growth (\%)}\\
  \midrule
  C1 & Slow	    & 2,706 (58.2\%) &    19.8 \\
  C2 & Moderate	& 1,399 (30.0\%) &    63.9 \\
	C3 & Fast	    & 434    (9.3\%) &   218.6 \\
  C4 & Viral	  & 110    (2.3\%) & 1,317.2 \\
  \bottomrule
\end{tabular}
\end{table}

\begin{figure}[!ht]
  \centering
  \includegraphics[width=0.475\linewidth,trim={0 4em 0 4em},clip]{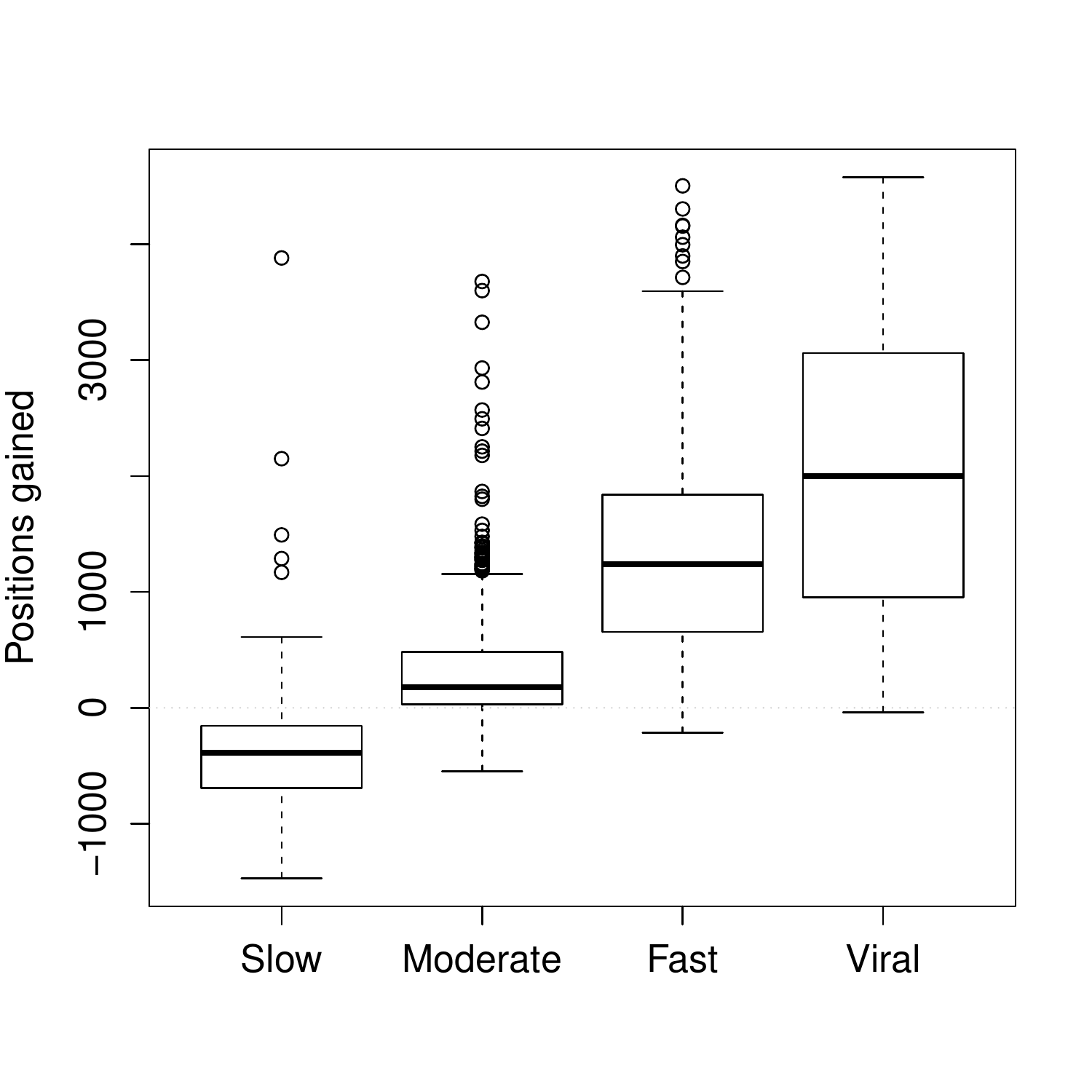}
  \caption{Rank differences in the interval of one year}
  \label{fig:patterns:ranking}
\end{figure}

We also investigate the correlation of the proposed growth patterns with the repositories ranking by number of stars.
To this purpose, we calculate the ranking of the studied repositories on week 0 (first week) and  51 (last week), by number of stars.
Next, we calculate the repositories rank in such weeks.
Repositories with positive values improved their ranking position, whereas negative values mean repositories losing positions.
Figure~\ref{fig:patterns:ranking} presents the distribution of the rank differences by growth pattern.
Initially, we can observe that at least 75\% of the slow repositories dropped in the ranking.
By contrast, almost all repositories (109 out of 110) with viral growth improved their rank in the same period.
Finally, 82\% and 96\% of the repositories with moderate and fast growth, respectively, increased their ranks.
By applying a Kruskal-Wallis test, we found that these distributions are different (\emph{p-value} $<$ 0.001).
According to Dunn's test, the rank differences of repositories with slow and moderate growth are statistically different from the other patterns; however, there is no statistical difference between repositories with fast and viral growth.

% !TEX root = jss.tex
\section{Growth Patterns Characterization}
\label{sec:factors}

In this section, we identify endogenous factors that distinguish the repositories in each growth pattern.
Revealing these factors is important because developers can strive to improve or change the ones that can be controlled or better understand the impact of those they have no control.

\subsection{Methodology}

To identify the most influential factors, we collected a set of characteristics of the repositories following each proposed growth pattern and applied a Random Forest classifier~\cite{breiman2001}.
We selected Random Forest because it is robust to noise and outliers~\cite{provost2001, Tian2015, andre2016}.

Table~\ref{tab:patterns:factors} lists 31 factors along three dimensions potentially affecting the stars growth of the repositories.
The \textsc{Repository} dimension includes factors that are accessible to users on the repositories' page in GitHub.
Usually, these pages are the main, or even unique, source of information about the projects and might influence the developers' decision on using (or not) a project.
For example, forks, and subscribers are measures of potential contributors to the repository.
Moreover, the quality of README files is another criterion considered by developers when selecting projects~\cite{begel2013}.

\begin{table}[!ht]
  \caption{Factors potentially affecting the growth pattern of a repository}
  \label{tab:patterns:factors}
  \resizebox{\textwidth}{!}{%
    \begin{tabular}{@{}cll@{}}
      \toprule
      \multicolumn{1}{c}{Dimension} & \multicolumn{1}{c}{Factor} & \multicolumn{1}{c}{Description} \\
      \midrule
      \multirow{16}{*}{Repository} & Stars (r.stars) & Number of stars \\
       & Forks (r.forks) & Number of forks \\
       & Network (r.network) & Number of repositories in the network\tablefootnote{Total number of forks including forks of forks.} \\
       & Subscribers (r.subscribers) & Number of users registered to receive notifications \\
       & Age (r.age) & Number of weeks since creation \\
       & Last Push (r.pushed) & Number of weeks since last \textit{git push} \\
       & Is Fork (r.is\textunderscore fork) & Repository is a fork (boolean value) \\
       & Has homepage (r.has\textunderscore homepage) & Repository has a homepage (boolean value) \\
       & Size (r.size) & Size of the repository in MB \\
       & Language (r.language) & Main programming language of the repository \\
       & Has Wiki (r.has\textunderscore wiki) & Repository has Wiki (boolean value) \\
       & Has Pages (r.has\textunderscore pages) & Repository has GitHub pages\tablefootnote{\url{https://pages.github.com}} (boolean value) \\
       & Is Mirror (r.mirror) & Repository is a mirror (boolean value) \\
       & Domain (r.domain) & Application domain (as defined in Section~\ref{sec:dataset}) \\
       & Description length (r.description\textunderscore length) & Number of words in the description \\
       & README length (r.readme\textunderscore length) & Number of words in the README file \\
       \midrule
       \multirow{9}{*}{Owner} & Account Type (o.type) & Account type: User or Organization\tablefootnote{\url{https://help.github.com/articles/what-s-the-difference-between-user-and-organization-accounts}} \\
       & Company (o.company) & Owner belongs to an organization (boolean value) \\
       & Has Public Email (o.email) & Owner has a public email (boolean value) \\
       & Public Repositories (o.repos) & Number of public repositories \\
       & Public Gists (o.gists) & Number of public code snippets \\
       & Followers (o.followers) & Number of followers \\
       & Following (o.followings) & Number of following \\
       & Total stars (o.stars) & Sum of all stars of all public repositories \\
       & Account Age (o.age) & Number of weeks since its account was created \\
       \midrule
       \multirow{6}{*}{\specialcell{Activity \\(last 52 weeks)}} & Commits (a.commits) & Number of commits \\
       & Contributors (a.contributors) & Number of contributors \\
       & Tags (a.tags) & Number of git tags \\
       & Releases (a.releases) & Number of releases \\
       & Issues (a.issues) & Number of issues \\
       & Pull Requests (a.pull\textunderscore requests) & Number of pull requests \\
      \bottomrule
    \end{tabular}
  }
\end{table}

The \textsc{Owner} dimension includes factors related to the repository' owner, for example, number of followers and account type.
For example, developers with more followers may take advantage of GitHub News Feed\footnote{\url{https://help.github.com/articles/news-feed}, a dashboard with recent activity on repositories.}, since their recent activities are shown to more developers~\cite{tsay2014}. Finally, developers owning popular repositories (by number of stars) might also attract more users to their other projects.

The \textsc{Activity} dimension includes factors related to the coding activity in the 52 weeks considered when extracting the growth patterns.
For example, higher number of commits might indicate that the project is in constant evolution whereas number of contributors, issues, and pull requests might indicate the engagement of the community with the project.

\begin{figure}[!ht]
  \centering
  \includegraphics[width=.775\linewidth,trim={0 0 0 0},clip]{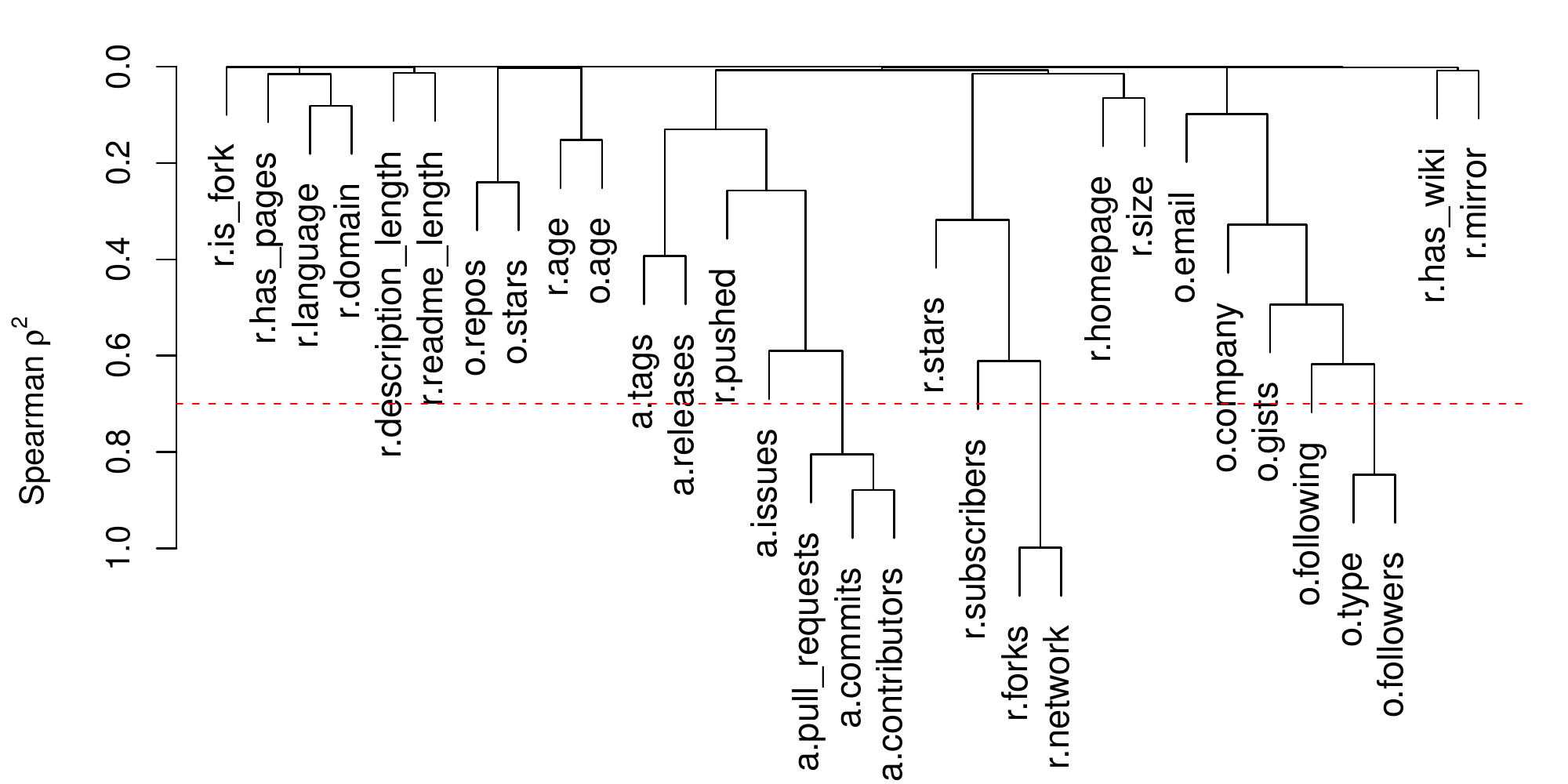}
  \caption{Correlation analysis (as result, we removed features \textit{a.pull\_requests}, \textit{a.contributors}, \textit{r.network}, and \textit{o.type})}
  \label{fig:classification:step1}
\end{figure}

Before using the Random Forest classifier, we performed a hierarchical cluster analysis on the 31 features in Table~\ref{tab:patterns:factors}.
This technique is proposed for assessing features collinearity and it is used in several other studies~\cite{Tian2015, RakhaSH16}.
Figure~\ref{fig:classification:step1} presents the final hierarchical cluster.
For sub-hierarchies with correlation greater than 0.7, only one variable was selected to the classifier.
For this reason, we removed the features \textit{a.pull\_requests} and \textit{a.contributors} (first cluster below the line), \textit{r.network} (second cluster), and \textit{o.type} (third cluster).

\subsection{Most Influential Factors}

To assess the relative importance of the selected features in discriminating each growth pattern, we used the \texttt{rfPermute} package for \texttt{R}~\cite{archer2013}.
We use the Mean Decrease Accuracy (MDA), which is determined during the prediction error measure phase, to rank the features based on their importance to the classifier.
MDA is quantified by measuring the change in prediction accuracy, when the values of the features are randomly permuted compared to the original observations~\cite{wolpert1999}.

Table~\ref{tab:patterns:factors:result} lists the top-10 most influential factors according to the feature importance ranking (all of them with \emph{p-value} $<$ 0.01).
As we can observe, these features are spread among the three dimensions, which shows their importance.
For \textsc{Repository}, the two most discriminative features are \textit{Age} and \textit{Last Push}, respectively.
In fact, for \textit{Age}, we observed that \textit{slow} growth is more common in old repositories whereas repositories presenting \textit{fast} and \textit{viral} growth are newest. The median number of weeks since creation is 235 for \textit{slow}, 167 for \textit{moderate}, 96 for \textit{fast}, and 76 for \textit{viral}.
Regarding \textit{Last Push}, we observed long inactive periods in repositories with \textit{slow} growth. The median number of weeks since the last code update is 3.53 for \textit{slow}, 0.80 for \textit{moderate}, 0.52 for \textit{fast}, and 0.49 for \textit{viral}.
For the \textsc{Owner} dimension, the two most discriminative features are \textit{Account Age} and \textit{Followers}, respectively.
The owners of repositories with \textit{viral} growth have the lowest account age (median of 173 weeks) and the lowest median number of followers (0).
Finally, for \textsc{Activity}, the two most discriminative features are \textit{Issues} and \textit{Commits}, respectively.
Similarly to previous factors, repositories with \textit{slow} growth have the lowest number of commits (only 19 commits).
Moreover, \textit{moderate} and \textit{fast} repositories have higher median number of issues than \textit{slow} and \textit{viral} repositories (51, 64, 19, and 11 issues, respectively).

\begin{table}[!ht]
  \caption{Top-10 most influential factors (\emph{p-value} $<$ 0.01)}
  \label{tab:patterns:factors:result}
  \centering
  \begin{tabular}{@{}cllc@{}}
    \toprule
    \multicolumn{1}{c}{\specialcell{Ranking}} & \multicolumn{1}{c}{Factor} & \multicolumn{1}{c}{Dimension} & \multicolumn{1}{c}{Actionable} \\
    \midrule
     1 & Age (r.age) & Repository & - \\
     2 & Last Push (r.pushed) & Repository & Yes \\
     3 & Issues (a.issues) & Activity & - \\
     4 & Commits (a.commits) & Activity & Yes \\
     5 & Forks (r.forks) & Repository & - \\
     6 & Account Age (o.age) & Owner & - \\
     7 & Stars (r.stars) & Repository & - \\
     8 & Subscribers (r.subscribers) & Repository & - \\
     9 & Followers (o.followers) & Owner & - \\
     10 & Tags (a.tags) &  Repository & Yes \\
    \bottomrule
  \end{tabular}
\end{table}

Although some factors cannot be controlled by developers, others depend on their actions.
From the top-10 most influential factors in Table~\ref{tab:patterns:factors:result}, three are directly impacted by developers' actions (column \aspas{Actionable}).
These results suggest that projects with frequent updates (\textit{Last Push}), a rich development history (\textit{Commits}), and frequent releases (\textit{Tags}) tend to attract more attention, in terms of number of stars.
However, it is also important to highlight that ``correlation does not necessarily imply in causation''. Therefore, it might be the project popularity that triggers constant pushes, commits, and releases. In other words, these results indicate that success in open source projects has its own price, which comes in the form of constantly having to update and improve the projects. Developers should be aware of this fact and reserve time to maintain a successful project. In fact, a recent survey shows that lack of time is the third most common reason for the failure of modern open source projects~\cite{fse2017}.

Finally, to assess the effectiveness of the classifier, we relied on metrics commonly used in Machine Learning and Information Retrieval~\cite{baeza1999}.
Precision measures the correctness of the classifier in predicting the repository growth pattern.
Recall measures the completeness of the classifier in predicting growth patterns.
F-measure is the harmonic mean of precision and recall.
Table~\ref{tab:patterns:confusion} shows the results for each growth pattern and the overall result.
In general, Random Forest performed satisfactorily for all patterns with a precision of 65.81\%, recall of 68.40\%, and F-measure of 67.08\%.
The \textit{Slow} pattern, which concentrates most of the repositories, presented the most accurate results (F-measure = 81.47\%).
On the other hand, Viral has the worst results (F-measure = 6.61\%), which can be caused by exogenous factors that are hard to predict.

\begin{table}[!ht]
  \caption{Classification effectiveness}
  \label{tab:patterns:confusion}
  \centering
  \begin{tabular}{@{}l|rrlrlrl@{}}
    \toprule
    Growth Pattern & & \multicolumn{2}{c}{Precision (\%)} & \multicolumn{2}{c}{Recall (\%)} & \multicolumn{2}{c}{F-measure (\%)} \\
    \midrule
    Slow     & & 75.98 & \sbarf{7598}{10000} & 87.80 & \sbarf{8780}{10000} & 81.47 & \sbarf{8147}{10000} \\
    Moderate & & 54.16 & \sbarf{5416}{10000} & 47.96 & \sbarf{4796}{10000} & 50.87 & \sbarf{5087}{10000} \\
    Fast     & & 47.43 & \sbarf{4743}{10000} & 29.72 & \sbarf{2972}{10000} & 36.54 & \sbarf{3654}{10000} \\
    Viral    & & 36.36 & \sbarf{3636}{10000} & 3.64 & \sbarf{364}{10000} & 6.61 & \sbarf{661}{10000}  \\
    \midrule
    Overall  & & 65.81 & \sbarf{6581}{10000} & 68.40 & \sbarf{6840}{10000} & 67.08 & \sbarf{6708}{10000} \\
    \bottomrule
  \end{tabular}
\end{table}

\vspace{0.5em}\begin{tcolorbox}[left=0.75em,right=0.75em,top=0.75em,bottom=0.75em,boxrule=0.25mm,colback=gray!5!white]
\noindent{\em Summary:} When we compare the proposed growth patterns, \textit{Age} is the most discriminative feature, followed by number of \textit{Issues} and \textit{Last Push}.
Moreover, three out of four features from the \textsc{Activity} dimension are in the top-10 most discriminative ones, which confirms the importance of constantly maintaining and evolving open source projects.
\end{tcolorbox}\vspace{0.5em}

% !TEX root =jss.tex
\section{Developers' Perceptions on Growth Patterns}
\label{sec:patterns-survey}

In this section, we describe a survey with developers to reveal their perceptions on the growth patterns proposed in this work.
Section~\ref{sec:patterns-survey:design} describes the design of the survey questionnaire and the selection of the survey participants.
Section~\ref{sec:patterns-survey:results} reports the survey results.

\subsection{Survey Design}
\label{sec:patterns-survey:design}

In this second survey, we asked developers to explain the reasons for the \textit{slow}, \textit{moderate}, \textit{fast}, or \textit{viral} growth observed in the number of stars of their repositories.
The questionnaire was sent by email to the repository's owner, for repositories owned by \textit{Users}, or to the contributor with the highest number of commits, for repositories owned by \textit{Organizations}.
For each growth pattern, we randomly selected 100 repositories whose developers have a public email.
Exceptionally for repositories classified with \textit{viral} growth, we selected 45 developers because they are the only ones with public emails on GitHub.
Thus, our sample of participants consists of 345 developers.

The questionnaire was sent between the 18th to 22nd of May 2017.
After a period of seven days, we received 115 responses, resulting in a response ratio of 33.3\%, considering the four growth patterns together (see details in Table~\ref{tab:patterns-survey:ratio}).
To preserve the respondents privacy, we use labels R1 to R115 when quoting their answers.
After receiving the answers, the first paper's author analyzed them, following the same steps of the survey presented in Section~\ref{sec:survey}.

\begin{table}[!ht]
\centering
\caption{Number of survey participants and answers per growth pattern\\CI = Confidence interval at confidence level of 95\%}
\label{tab:patterns-survey:ratio}
\begin{tabular}{@{}lcccc@{}}
  \toprule
  Growth Pattern & \multicolumn{1}{c}{Participants} & \multicolumn{1}{c}{Answers} & \multicolumn{1}{c}{\%} & \multicolumn{1}{c}{CI} \\
  \midrule
  Slow & 100 & 26 & 26.0 & 19.1  \\
  Moderate & 100 & 33 & 33.0 & 16.9 \\
  Fast & 100 & 34 & 34.0 & 16.1  \\
  Viral & 45 & 22 & 48.9 & 18.8  \\
  \bottomrule
\end{tabular}
\end{table}

\subsection{Survey Results}
\label{sec:patterns-survey:results}

Table~\ref{tab:patterns-survey:slow} lists five major reasons for \textit{slow} growth, according to the surveyed developers.
Unmaintained or low activity was the main reason, reported by 14 developers (53.8\%).
Limited or lack of promotion was mentioned by four developers (15.3\%).
For three developers, the project focus on a specific niche audience, thus not being so popular as other repositories.
Furthermore, emergence of alternative solutions was the reason pointed by two developers.
Other three developers reported they have no idea on the reasons of the \textit{slow} growth.
Finally, five developers provided other reasons (e.g., project age).
Examples of answers include:\medskip

\noindent{\textit{The reason is there's no new material there. Also the material that is there is becoming outdated and less relevant over time.}} (R27, Unmaintained or low activity)\medskip

\noindent{\textit{I believe the primary reason is that I am doing virtually nothing to actively promote the project.}} (R26, Limited or lack of promotion)\medskip

\noindent{\textit{I don’t know the root cause, my guess is that it’s a rather specialized tool with a limited audience.}} (R38, Niche audience)\medskip

\begin{table}[!ht]
\centering
\caption{Reasons for Slow Growth\\(95\% confidence level with a 19.1\% confidence interval)}
\label{tab:patterns-survey:slow}
\begin{tabular}{@{}lcc@{}}
\toprule
Reason & Answers & \multicolumn{1}{c}{Percentage (\%)}  \\
\midrule
Unmaintained or low activity & 14 & 53.8 \sbarf{14}{26}  \\
Limited or lack of promotion & 4 & 15.3 \sbarf{4}{26}  \\
Niche audience & 3 & 11.5 \sbarf{3}{26}  \\
Alternative solutions & 2 & \enspace7.6 \sbarf{2}{26}  \\
Unknown & 3 & 11.5 \sbarf{3}{26}  \\
Other reasons & 5 & 19.2 \sbarf{5}{26}  \\
\bottomrule
\end{tabular}
\end{table}

After analyzing the reasons for \textit{moderate} growth, we identified two conflicting sentiments in the answers: (a) \textit{positives} reasons, which are contributing to the stars growth; (b) \textit{negative} reasons, which are limiting the stars growth.
Table~\ref{tab:patterns-survey:moderate} lists the major reasons for the \textit{positive} and \textit{negative} sentiments.

For \textit{positive} sentiments, 15 developers (45.4\%) mentioned active promotion (mainly on social media sites, as Hacker News\footnote{\url{https://news.ycombinator.com}}).
The use of trending technologies was mentioned by nine developers (27.2\%).
For example, \textsc{danialfarid/ng-file-upload} (a popular \textsc{Angular} component) is benefited by the large community of \textsc{Angular} practitioners.
Active project (e.g., with frequent updates and fast issues resolution) was mentioned by seven developers (21.2\%).
Three developers explicitly mentioned the repository provides an innovative solution and two developers mentioned that code or documentation quality contributed to the stars growth.
Finally, three other positive reasons were provided (project usability, usefulness, and maturity).
As examples we have these positive answers:\medskip

\noindent{\textit{It could be related to how many people are using Angular JS and the development and new features in the module had been active for couple years.}} (R34, Trending technology, Active project)\medskip

\noindent{\textit{The initial increase in stars happened as word of the project got out. I initially had a Product Hunt page and posted it on Hacker News. From there it is started to popup on other tech sites.}} (R85, Active promotion)\medskip

\noindent{\textit{Our continued releases every 3-4 months for nearly 6 years is probably the reasoning. We are a steady, stable, open source solution for reverse engineering.}} (R16, Active project, Maturity)\medskip

\begin{table}[!ht]
\centering
\caption{Reasons for Moderate Growth\\(95\% confidence level with a 16.9\% confidence interval)}
\label{tab:patterns-survey:moderate}
\resizebox{\textwidth}{!}{%
  \begin{tabular}{@{}lcc|lcc@{}}
  \toprule
  \multicolumn{3}{c|}{Positive Sentiments} & \multicolumn{3}{c}{Negative Sentiments} \\
  \midrule
  Reason & Answers & \multicolumn{1}{c|}{Percentage (\%)} & Reason & Answers & \multicolumn{1}{c}{Percentage (\%)} \\
  \midrule
  Active promotion & 15 & 45.4 \sbarf{15}{33} & Niche audience  & 3 & \enspace9.0 \sbarf{3}{33}  \\
  Trending technology & 9 & 27.2 \sbarf{9}{33} & Low activity & 2 & \enspace6.0 \sbarf{2}{33} \\
  Active project & 7 & 21.2 \sbarf{7}{33} & Limited or lack of promotion & 1 & \enspace3.0 \sbarf{1}{33} \\
  Innovative project & 3 & \enspace9.0 \sbarf{3}{33} & Old project & 1 & \enspace3.0 \sbarf{1}{33} \\
  Code or doc. quality & 2 & \enspace6.0 \sbarf{2}{33} &  \\
  Other & 3 & \enspace9.0 \sbarf{3}{33} &  &  & \\
  \bottomrule
  \end{tabular}
}
\end{table}

For answers transmitting \textit{negative} sentiments, three developers mentioned the project's niche audience as a restrictive growth factor.
Moreover, low activity and limited or lack of promotion were mentioned by two and one developers, respectively.
Finally, one developer mentioned that the project age is restricting its stars growth.
Examples of negative answers are:\medskip

\noindent{\textit{I think the demographics for [repository] users shifts towards the [other-repository] -- new devs and people new to a young language tend to look for more features, and [repository] is explicitly not that.}} (R25, Niche audience)\medskip

\noindent{\textit{My best guess is that it's an older project that's occasionally attracting new people, but there's no single big \aspas{marketing event} where it gets a huge spike of GitHub stars.}} (R28, Old project, Limited or lack of promotion)\medskip

For repositories presenting \textit{fast} growth, Table~\ref{tab:patterns-survey:fast} lists six major reasons reported by their developers.
Active promotion is the major reason according to 22 developers (64.7\%).
Furthermore, trending technology was mentioned by 11 developers (32.3\%).
Other eight developers (24.5\%) mentioned that it is an innovative project.
Examples of reasons for \textit{fast} growth include:\medskip

\noindent{\textit{It's a popular project because nothing else like it exists for React.}} (R72, Innovative project, Trending technology)\medskip

\noindent{\textit{We've been adding a lot of features in the last year, and I've been trying to evangelise the project to gain new users - some of those things probably helped a lot.}} (R66, Active project, Active promotion)\medskip

\begin{table}[!ht]
\centering
\caption{Reasons for Fast Growth\\(95\% confidence level with a 16.1\% confidence interval)}
\label{tab:patterns-survey:fast}
\begin{tabular}{@{}lcc@{}}
\toprule
Reason & Answers & \multicolumn{1}{c}{Percentage (\%)}  \\
\midrule
Active promotion & 22 & 64.7 \sbarf{22}{34}  \\
Trending technology & 11 & 32.3 \sbarf{11}{34}  \\
Innovative project & 8 & 24.5 \sbarf{8}{34}  \\
Active project & 5 & 14.7 \sbarf{5}{34}  \\
Project usability & 2 & \enspace5.8 \sbarf{2}{34}  \\
Project usefulness & 2 & \enspace5.8 \sbarf{2}{34}  \\
Unknown & 2 & \enspace5.8 \sbarf{2}{34}  \\
Other & 5 & 14.7 \sbarf{2}{34}  \\
\bottomrule
\end{tabular}
\end{table}

Finally, Table~\ref{tab:patterns-survey:viral} lists five major reasons that emerged after analysing the developers' answers for \textit{viral} growth.
As observed, 16 developers (72.7\%) linked this behavior to successful posts in social media sites, mostly Hacker News.
Code or documentation quality were mentioned by six developers (27.2\%).
Four developers (19.0\%) linked the \textit{viral} growth to trending technologies.
As examples of answers we have:\medskip

\noindent{\textit{Yes, we had a huge bump in stars. The secret: coverage by Hacker News, which resulted in follow-up by other news sites.}} (R44, Promotion on social media sites)\medskip

\noindent{\textit{In my opinion is just that [repository] replied to some people need and gain adoption very fast. Sharing the project on reddit/twitter/hacker news helped a lot the spread of it. In my opinion the quality of docs/examples helps a lot.}} (R103, Promotion on social media sites, Code or documentation quality, Useful project)\medskip

\noindent{\textit{I believe the project has seen such great growth because of it's position within the greater Angular community ...}} (R87, Trending technology)\medskip

\begin{table}[!ht]
\centering
\caption{Reasons for Viral Growth\\(95\% confidence level with a 18.8\% confidence interval)}
\label{tab:patterns-survey:viral}
\begin{tabular}{@{}lcc@{}}
\toprule
Reason & Answers & \multicolumn{1}{c}{Percentage (\%)}  \\
\midrule
Promotion on social media sites & 16 & 72.7 \sbarf{16}{21}  \\
Code or documentation quality & 6  & 27.2 \sbarf{6}{21}  \\
Trending technology & 4  & 19.0 \sbarf{4}{21}  \\
Useful & 3  & 14.2 \sbarf{3}{21}  \\
New features & 2  & \enspace9.5 \sbarf{2}{21}  \\
Other & 2  & \enspace9.5 \sbarf{2}{21}  \\
Unknown & 1  & \enspace4.7 \sbarf{1}{21}  \\
\bottomrule
\end{tabular}
\end{table}

\vspace{0.5em}\begin{tcolorbox}[left=0.75em,right=0.75em,top=0.75em,bottom=0.75em,boxrule=0.25mm,colback=gray!5!white]
\noindent{\em Summary:}
According to the surveyed developers, the major reason for \textit{slow} growth is deprecation or lack of activity (53.8\%).
Regarding \textit{moderate} growth, there are two conflicting sentiments on the developers' answers: positive sentiments (e.g., active promotion) and negative sentiments (e.g., niche audience).
For \textit{fast} growth, the three major reasons are active promotion, usage of trending technology, and innovative project.
Finally, the major reason for \textit{viral} growth is also promotion on social media sites (72.7\%). 
\end{tcolorbox}\vspace{0.5em}

\noindent {\em Implications for Empirical Software Engineering Researchers:} The following observations are derived in this second survey regarding the selection of GitHub projects based on number of stars: (1) this selection might favor projects with successful marketing and advertising strategies, despite the adoption of well-established software engineering practices; (2) it is particularly important to check whether the projects have a viral growth behavior (e.g., {\sc chrislgarry/Apollo--11} gained  19,270 stars in just two weeks).
% !TEX root = jss.tex
\section{Threats to Validity}
\label{sec:threats}

\noindent{\em Dataset.} GitHub has millions of repositories.
We build our dataset by collecting the top-5,000 repositories by number of stars, which represents a small fraction in comparison to the GitHub's universe.
However, our goal is exactly to investigate the most starred repositories.
Furthermore, most GitHub repositories are forks and have very low activity~\cite{Kalliamvakou2014, Kalliamvakou2015, cosentino2017}.\bigskip

\noindent{\em Application domains.} Because GitHub does not classify the repositories in domains, we performed this classification manually. Therefore, it is subjected to errors and inaccuracies. To mitigate this threat, the dubious classification decisions were discussed by the two paper's authors.\bigskip

\noindent{\em Survey study 1.} The 5,000 repositories in our dataset have more than 21 million stars together.
Despite this fact, we surveyed only the last developers who starred these repositories, a total of 4,370 developers.
This decision was made to do not spam the developers.
Moreover, we restricted the participants to those who gave a star in the last six months to increase the chances they remember the motivation for starring the projects.
Another threat is related to the manual classification of the answers to derive the starring motivations.
Although this activity has been done with special attention by the paper's first author, it is subjective by nature.\bigskip

\noindent{\em Survey study 2.} In the second survey, we asked the developers to explain the reasons for the \textit{slow}, \textit{moderate}, \textit{fast}, or \textit{viral} growth observed in the number of stars of their repositories.
For each growth pattern, we randomly selected a group of 100 repositories/developers.
Exceptionally for repositories presenting a \textit{viral} growth, 45 developers were used since they are the only ones with public e-mails.
Since we received 115 answers (corresponding to a response ratio of 33.3\%), we report the perceptions of a non-negligible number of developers.\bigskip

\noindent{\em Growth patterns}. The selection of the number of clusters is a key parameter in algorithms like KSC. To mitigate this threat, we employed a heuristic that considers the intra/intercluster distance variations~\cite{Menasce2001}. Furthermore, the analysis of growth patterns was based on the stars obtained in the last year. The stars before this period are not considered, since KSC requires time series with the same length.\bigskip

\noindent{\em Growth patterns characterization}. In Section~\ref{sec:factors}, we use a random forest classifier to identify the factors that distinguish the proposed growth patterns. This classifier requires the number of trees to compose a Random Forest.
In this study, we used 100 trees, which is in the range suggested by Oshiro et. al.~\cite{oshiro2012}.\bigskip

% !TEX root = jss.tex
\section{Related Work}
\label{sec:related-work}

We organize related work in four groups: (1) criteria for selecting GitHub projects; (2) studies on GitHub popularity metrics; (3) popularity of mobile apps; and (4) popularity of social media content.\medskip

\noindent\textbf{Criteria for Selecting GitHub Projects:}
Stars are often used by researchers to select GitHub projects for empirical studies in software engineering~\cite{ray2014, padhye2014study, hilton2016usage, MazinanianKTD17, JIANG201744, Nielebock2018,Rigger2018, castro2018}.
For example, in a previous study, we use the top-5,000 GitHub repositories with most stars to investigate the performance of linear regression models to predict the number of stars in the future~\cite{borges2016_2}.
In a more recent study, we use the top-100 most starred GitHub repositories to investigate the channels used by open source project managers to promote their systems~\cite{borges2018}.
Ray et al.~select 50 projects by stars on GitHub to study the effects of programming language features on defects~\cite{ray2014}.
To study the levels of participation of different open-source communities, Padhye et al.~rely on the 89 most-starred GitHub projects~\cite{padhye2014study}.
Hilton et al.~study Continuous Integration (CI) practices using a sample of 50 projects, ranked by number of stars~\cite{hilton2016usage}.
Silva et al.~select 748 Java projects to study refactoring practices among GitHub contributors~\cite{danilo2016} and Mazinanian study the adoption of lamba expressions in a large sample of 2,000 Java projects, also ordered by stars~\cite{MazinanianKTD17}.
Finally, Castro and Schots use GitHub stars as cut-off criterion to select projects and then analyze logging information to propose a visualization tool~\cite{castro2018}.

However, there are also studies that rely on different metrics and methodologies to select GitHub projects.
For example, Vasilescu et al.~use the programming language and number of forks to collect 246 repositories and then characterize the effects of CI in process automation on open source projects~\cite{Vasilescu2015}.
Bissyandé et al.~use the default criteria of the GitHub API (i.e., best match) to collect 100K repositories and study popularity, interoperability, and impact of programming languages~\cite{Bissyande2013}.
Kikas et al.~combine number of issues and commits to obtain a large set of GitHub projects and study models to predict whether an issue will be closed~\cite{Kikas:2016}.

Finally, there are efforts proposing more rigorous methods to select projecs in software repositories. For example, 
Falessi et al.~first perform a systematic mapping study with 68 past studies and did not find any study that can be ranked as completely replicable~\cite{Falessi2017}. Then, the authors present a rigorous method to select projects, called STRESS, that allows users to define the desired level of diversity, fit, and quality. Munaiah et al.~propose a similar framework and tool, called Reaper, to select engineering GitHub projects, i.e., projects that follow sound software engineering practices, including  documentation, testing, and project management~\cite{MunaiahKCN17}. Ultimately, Reaper was conceived to to separate the signal (e.g.,~engineering  software projects) from the noise (e.g.,~home work assignments) when selecting projects in GitHub.
As part of their findings, the authors report that using stars to classify engineered GitHub projects results on a very high precision, but with a low recall. In other words, repositories with a large number of stars are usually engineered projects; however, the contrary is not always true. Previously, Nagappan et al. proposed a measure, called sample coverage, to capture the percentage of projects in a population that are similar to a given sample~\cite{Nagappan2013FSE}. Their goal is to promote the importance of diversity when selecting projects for evaluating a software engineering approach or performing an empirical study. They illustrated the usage of
sample coverage in a population of 20K projects monitored by Ohloh.net, which is a public directory of open source projects, currently called Open Hub.\medskip

\noindent\textbf{Studies on GitHub Popularity Metrics:}
Several studies investigate characteristics and usages of GitHub popularity metrics.
Zho et al.~study the frequency of folders used by 140K GitHub projects and their results suggest that the use of standard folders (e.g., doc, test, examples) may have an impact on project popularity, in terms of number of forks~\cite{Zhu2014}.
Aggarwal et al.~study the effect of social interactions on GitHub projects' documentation~\cite{Aggarwal2014}. They conclude that popular projects tend to attract more documentation collaborators.
Jiang et al.~provide a comprehensive analysis of inactive yet available assignees in popular GitHub projects. They show that some projects have more than 80\% of inactive assignees~\cite{JIANG201744}.
Wanwangying et al.~conduct a study to identify the most influential Python projects on GitHub~\cite{Ma2016}. They found that the most influential projects are not necessarily popular among GitHub users.
By analyzing the effect of evolutionary software requirements on open source projects, Vlas et al.~state that popularity (measured by number of stars and forks) depends on the continuous developing of requirements~\cite{vlas2017}.
Papamichail et al.~argue that the popularity of software components is as an indicator of software quality~\cite{Papamichail2016}; however, Herraiz et al.~alert that popularity can also impact the perceived quality~\cite{debian_wcre2011}.
Finally, as one of the findings of a systematic mapping study, Cosentino et al.~report that popularity (as measured by number of stars) is also useful to attract new developers to open source projects~\cite{cosentino2017}.\medskip

\noindent\textbf{Popularity of mobile apps:} Popularity in the context of mobile apps is the subject of several studies. For example, there are many studies examining the relationship between popularity of mobile apps and code properties~\cite{Fu2013, LinaresVasquez2013, Ruiz2014, lee2014, Tian2015, Corral2015, McIlroy2016}.
Yuan et al.~investigate 28 factors along eight dimensions to understand how high-rated Android applications are different from low-rated ones~\cite{Tian2015}. Their results show that external factors, like number of promotional images, are the most influential ones.
Guerrouj and Baysal explore the relationships between mobile apps' success and API quality~\cite{Guerrouj2016}. They found that changes and bugs in API methods are not strong predictors of apps' popularity.
McIlroy et al.~study the frequency of updates in popular free apps from different categories in the Google Play store~\cite{McIlroy2016}. They report that frequently-updated apps do not experience an increase in negative ratings by their users.
Ruiz et al.~examine the relationship between the number of ad libraries and app's user ratings~\cite{Ruiz2014}. Their results show that there is no relationship between these variables.
Lee and Raghu tracked popular apps in the Apple Store and found that the survival rates of free apps are up to two times greater than the paid ones~\cite{lee2014}. Moreover, they report that frequent feature updates can contribute to app survival among the top ones.
Ali et al.~conducted a comparative study of cross-platform apps to understand their characteristics~\cite{Ali2017}. They show that users can perceive and rate differently the same app on different platforms.\medskip

\noindent\textbf{Popularity of social media content:}
Other studies track popularity on social networks, including video sharing sites (e.g., YouTube) and social platforms (e.g., Twitter and news aggregators).
Chatzopoulou et al.~\cite{Chatzopoulou2010} analyze popularity of YouTube videos by looking at properties and patterns metrics. They report that several popularity metrics are highly correlated.
Lehmann et al.~\cite{Lehmann2012} analyze popularity peaks of hashtags. They found four usage patterns restricted to a two-week period centered on the peak time.
Aniche et al.~conduct a study to understand how developers use modern news aggregator sites (Reddit and Hacker News)~\cite{aniche2018}. According to their results, half of the participants read only the most upvoted comments and posts.\medskip
% !TEX root = jss.tex
\section{Conclusion}
\label{sec:conclusion}

In this paper, we reported that developers star GitHub repositories due to three major reasons (which frequently overlap): to show appreciation to projects, to bookmark a project, and because they are using a project. Furthermore, three out of four developers declared they consider the number of stars before using or contributing to GitHub projects.

\vspace{0.5em}\begin{tcolorbox}[left=0.75em,right=0.75em,top=0.75em,bottom=0.75em,boxrule=0.25mm,colback=gray!5!white]
\noindent{\em Recommendation \#1:}
Stars are a key metric about the evolution of GitHub projects; therefore, project managers should track and compare the number of stars of their projects with competitor ones.
\end{tcolorbox}\vspace{0.5em}

We provided a quantitative characterization of the top-5,000 most  starred repositories.
We found that repositories owned by organizations have more stars than the ones owned by individuals (RQ \#1).
We also reported the existence of a moderate correlation of stars with contributors and forks, a low correlation between stars and commits, and no correlation between stars and repository' age (RQ \#2).
Furthermore, repositories have a tendency to receive more stars right after their public release (RQ \#3).
Finally, there is an acceleration in the number of stars gained after releases (RQ \#4).

%\vspace{0.5em}\begin{tcolorbox}[left=0.75em,right=0.75em,top=0.75em,bottom=0.75em,boxrule=0.25mm,colback=gray!5!white]
%\noindent{\em Recommendation \#2:}
%Project managers should consider using organizational accounts (e.g., {\em aserg-ufmg} instead of {\em hsborges}). It is also important to work to attract new contributors and to evolve the projects by frequently providing new releases.
%\end{tcolorbox}\vspace{0.5em}

We validated the proposed stars growth patterns by means of a survey with project owners and core developers. We revealed that the major reason for a \textit{slow} growth in the number of stars is project deprecation or inactivity.
Regarding \textit{moderate} growth, we detected both positive sentiments (e.g., active promotion) and negative ones (e.g., niche audience).
The major reasons for \textit{fast} growth are active promotion, usage of trending technologies, and innovative projects.
Finally, the major reason for \textit{viral} growth is also promotion on social media.

\vspace{0.5em}\begin{tcolorbox}[left=0.75em,right=0.75em,top=0.75em,bottom=0.75em,boxrule=0.25mm,colback=gray!5!white]
\noindent{\em Recommendation \#2:}
Open source projects require an investment on marketing and advertisement, mainly in social networks and programming forums, like Hacker News.
\end{tcolorbox}\vspace{0.5em}

We distilled a list of threats practitioners and researchers may face when selecting GitHub projects based 
on the number of stars. For example, this selection favors large projects, with many contributors and forks. It may also include projects that receive a large number of stars in a short interval, including projects with a viral growth in their number of stars. Finally, it tends to favor projects with effective marketing and advertising strategies, which do not necessarily follow solid software engineering principles and practices.

\vspace{0.5em}\begin{tcolorbox}[left=0.75em,right=0.75em,top=0.75em,bottom=0.75em,boxrule=0.25mm,colback=gray!5!white]
\noindent{\em Recommendation \#3:}
When selecting projects by number of stars, practitioners and researchers should check whether the stars are not concentrated in a short time period or whether they are mostly a consequence of active promotion in social media sites.
\end{tcolorbox}\vspace{0.5em}

Future work may include an investigation of repositories that have few stars, including a comparison with the most starred ones. It would also be interesting to correlate repository's stars and language popularity and in this way to investigate relative measures of popularity. For example, if we restrict the analysis to a given language, a Scala repository can be considered more popular than a JavaScript one, although having less stars. Finally, the use of a different technique (e.g.,~Scott-Knott ESD~\cite{tantithamthavorn2017mvt}) may provide additional insights on the factors that impact the classification of a project in a given growth pattern.

\vspace{0.5em}\begin{tcolorbox}[left=0.75em,right=0.75em,top=0.75em,bottom=0.75em,boxrule=0.25mm,colback=gray!5!white]
\noindent{\em Tool and dataset:}
We implemented a tool to explore and check our results, including the time series of stars used in this paper and the proposed growth patterns. It is available at: \url{http://gittrends.io}.
The analyzed data, manual classification of the application domain, and the surveyed responses used in this study are publicly available at: \url{https://doi.org/10.5281/zenodo.1183752}.
\end{tcolorbox}\vspace{0.5em}

\section*{Acknowledgments}

\noindent This research is supported by CAPES, CNPq, and FAPEMIG.

\singlespacing

\section*{Bibliography}
\bibliographystyle{elsarticle-num}
\bibliography{references}

\end{document}